\documentclass[10pt,twocolumn,letterpaper]{article}
\usepackage[utf8]{inputenc} %
\usepackage[T1]{fontenc}    %
\usepackage{xcolor}         %
\usepackage{url}            %
\usepackage{booktabs}       %
\usepackage{amsfonts}       %
\usepackage{nicefrac}       %
\usepackage{microtype}      %

\usepackage{mathtools}
\usepackage{amsmath}

\newcommand\eat[1]{}
\usepackage{graphicx}
\usepackage{subcaption}

\DeclarePairedDelimiter\round{\lfloor}{\rceil}

\newcommand{\citep}[1]{\cite[#1]} %
\usepackage{multirow}
\usepackage{makecell}

\def\E{{\mathbb E}}
\def\setX{{\mathcal{X}}}

\def\sethX{\hat{\mathcal{X}}}
\def\setZ{{\mathcal{Z}}}
\def\a{\mathbf{a}}

\def\x{{\mathbf x}}

\def\z{{\mathbf z}}
\def\br{{\mathbf r}}

\def\h{{\mathbf h}}
\def\X{{\mathbf X}}

\def\hatx{{\mathbf{\hat x}}}

\def\hatz{{\mathbf{\hat z}}}

\usepackage{lipsum}  %
\newcommand\blfootnote[1]{%
  \begingroup
  \renewcommand\thefootnote{}\footnote{#1}%
  \addtocounter{footnote}{-1}%
  \endgroup
}

\usepackage{iccv}
\usepackage{times}
\usepackage{epsfig}
\usepackage{graphicx}
\usepackage{amsmath}
\usepackage{amssymb}
\usepackage{subcaption}
\usepackage[accsupp]{axessibility}  %

\usepackage[pagebackref=true,breaklinks=true,letterpaper=true,colorlinks,bookmarks=false]{hyperref}

\iccvfinalcopy %

\def\iccvPaperID{12730} %

\ificcvfinal\fi

\begin{document}

\title{
Computationally-Efficient Neural Image Compression with Shallow Decoders
}

\author{Yibo Yang \qquad \qquad 
 Stephan Mandt\\
Department of Computer Science\\
University of California, Irvine \\
{\tt\small \{yibo.yang, mandt\}@uci.edu}
}

\maketitle
\ificcvfinal\thispagestyle{empty}\fi

\begin{abstract}

Neural image compression methods have seen increasingly strong performance in recent years. However, they suffer orders of magnitude higher computational complexity compared to traditional codecs, which hinders their real-world deployment.
This paper takes a step forward towards closing this gap in decoding complexity by using a shallow or even linear decoding transform resembling that of JPEG.
To compensate for the resulting drop in compression performance, we exploit the often asymmetrical computation budget between encoding and decoding, by adopting more powerful encoder networks and iterative encoding. We theoretically formalize the intuition behind, and our experimental results establish a new frontier in the trade-off between rate-distortion and decoding complexity for neural image compression. Specifically, we achieve rate-distortion performance competitive with the established mean-scale hyperprior architecture of Minnen et al. (2018) at less than 50K decoding FLOPs/pixel, reducing the baseline's overall decoding complexity by 80\%, or over 90\% for the synthesis transform alone.
Our code can be found at \url{https://github.com/mandt-lab/shallow-ntc}. %

\end{abstract}

\section{Introduction}

Deep-learning-based methods for data compression \cite{yang2022introduction} have achieved increasingly strong performance on visual data compression, increasingly exceeding classical codecs in rate-distortion performance \cite{balle2017end, minnen2018joint, yang2020variational, cheng2020learned, yang2020hierarchical, mentzer2022vct}.
However, their enormous computational complexity compared to classical codecs, especially required for decoding, is a roadblock towards their wider adoption \cite{minnen_current_2021, mukherjee_challenges_2022}. 
In this work, inspired by the parallel between nonlinear transform coding and traditional transform coding \cite{duan2022opening}, we replace deep convolutional decoders with extremely lightweight and shallow (and even linear) decoding transforms, and establish the R-D (rate-distortion) performance of neural image compression when operating at the lower limit of decoding complexity.

\begin{figure}[t]
    \centering
    \includegraphics[width=0.95\linewidth]{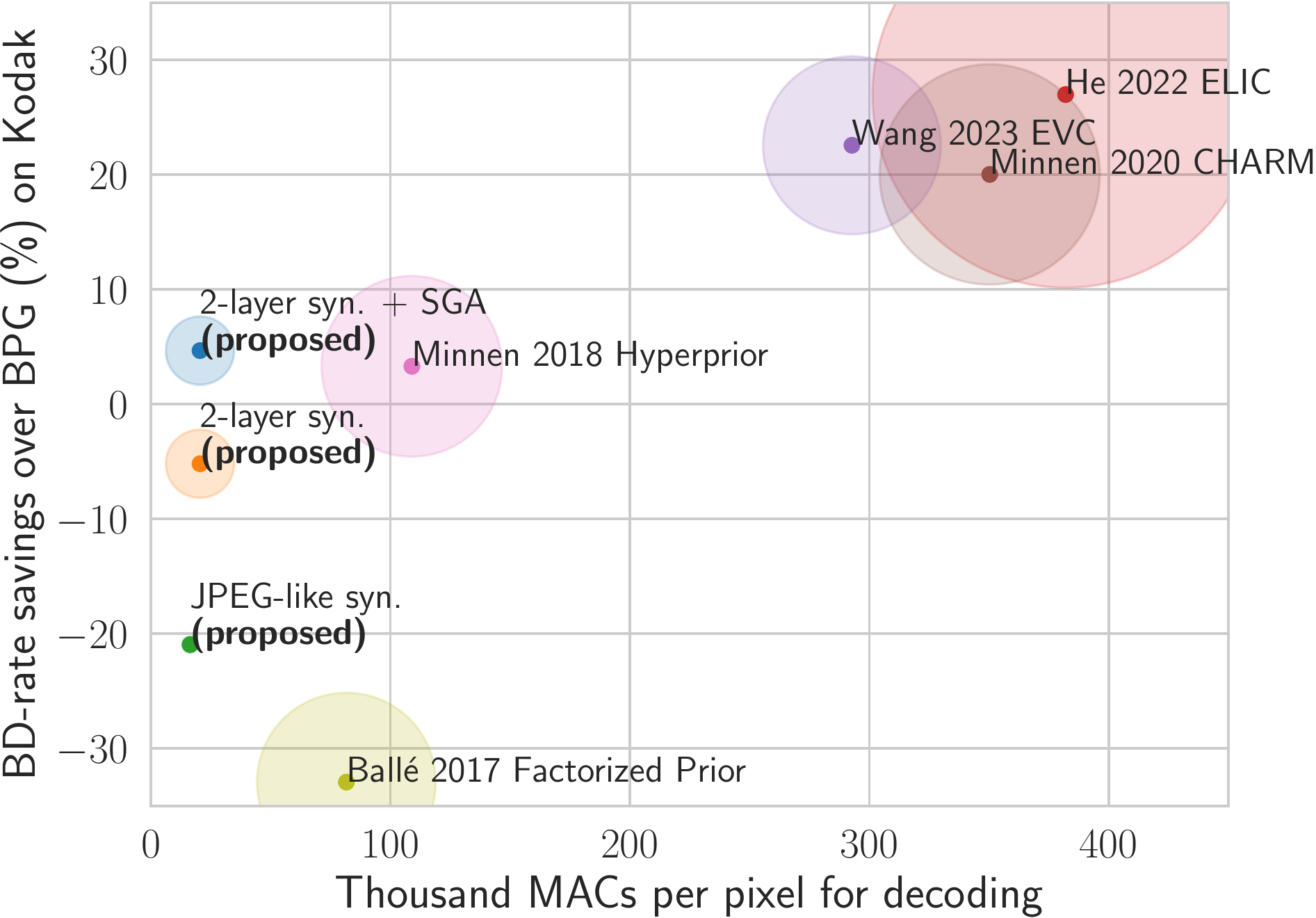}
    \caption{R-D performance on Kodak v.s. decoding computation complexity as measured in KMACs (thousand multiply-accumulate operations) per pixel.  The circle radius corresponds to the parameter count of the synthesis transform in each method (see Table.~\ref{tab:macs})
    }
    \label{fig:rdc_kodak}
\end{figure}

More concretely, our contributions are as follows:
\begin{itemize}
    \item We offer new insight into the image manifold parameterized by learned synthesis transforms in nonlinear transform coding. Our results suggest that the learned manifold is relatively flat and preserves linear combinations in the latent space,
    in contrast to its highly nonlinear counterpart in generative modeling \cite{chen2020learning}. 
    \item Inspired by the parallel between neural image compression and traditional transform coding, we study the effect of linear synthesis transform within a hyperprior architecture. We show that, perhaps surprisingly, a JPEG-like synthesis can perform similarly to a deep linear CNN, and we shed light on the role of nonlinearity in the perceptual quality of neural image compression.
    \item We give a theoretical analysis of the R-D cost of neural lossy compression in an asymptotic setting, which quantifies the performance implications of varying the complexity of encoding and decoding procedures.
    \item We equip our JPEG-like synthesis with powerful encoding methods, and augment it with a single hidden layer. This simple approach yields a new state-of-the-art result in the trade-off between R-D performance and decoding complexity for nonlinear transform coding, in the regime of sub-50K FLOPs per pixel  believed to be dominated by classical codecs. 
\end{itemize}

\section{Background and notation}
\subsection{Neural image compression} \label{sec:ntc}

Most existing neural lossy compression approaches are based on the paradigm of nonlinear transform coding (NTC) \cite{balle2021ntc}.
Traditional transform coding \cite{goyal2000transform} involves designing a pair of analysis (encoding) transform $f$ and synthesis (decoding) transform $g$ such that the encoded representation of the data achieves good R-D (rate-distortion) performance. NTC essentially learns this process through data-driven optimization.  
Let the input color image be $\x \in \mathbb{R}^{H \times W \times 3}$.
The analysis transform computes a continuous latent representation $\z := f(\x)$, which is then quantized to $\hatz = \round{\z}$ and transmitted to the receiver under an entropy model $P(\hatz)$; the final reconstruction is then computed by the synthesis transform as $\hatx := g(\hatz)$. The hard quantization is typically replaced by uniform noise to enable end-to-end training \cite{balle2016end}. We refer to \cite{yang2022introduction} (Section 3.3.3) for the technical details.

Instead of orthogonal linear transforms in traditional transform coding, the analysis and synthesis transforms in NTC are typically CNNs (convolutional neural networks) \cite{theis2017cae, balle2016end} or variants with residual connections or attention mechanisms \cite{cheng2020learned, he2022elic}. The (convolutional) latent coefficients $\z \in \mathbb{R}^{h, w, C}$  form a 3D tensor with $C$ channels and a spatial extent $(h, w)$ smaller than the input image. 
We denote the downsampling factor by $s$, i.e., $ s = H / h = W / w$; this is also the ``upsampling'' factor of the synthesis transform.

To improve the bitrate of NTC, a \emph{hyperprior} \cite{balle2018hyper, minnen2018joint} is commonly used to parameterize the entropy model $P(\hatz)$ via another set of latent coefficients $\h$ and an associated pair of transforms $(f_h, g_h)$. The hyper analysis $f_h$ computes $\h = f_h(\hatz)$ at encoding time, and the hyper synthesis $g_h$ predicts the (conditional) entropy model $P(\hatz | \hat{\h})$ based on the quantized $\hat{\h} = \round{\h}$. 
We adopt the Mean-scale Hyperprior  from Minnen et al. \cite{minnen2018joint} as our base architecture, which is widely used as a basis for other NTC methods \cite{johnston2019computationally, cheng2020learned, minnen2020channel, he2022elic}. In this architecture, the various transforms are parameteried by CNNs, with GDN activation \cite{balle2016end} being used in the analysis and synthesis transforms and ReLU activation in the hyper transforms.  Importantly, the synthesis  transform ($g$) accounts for over 80\% of the overall decoding complexity (see Table~\ref{tab:macs}), and is the focus of this work.

\subsection{Iterative inference}
Given an image $\x$ to be encoded, instead of computing its discrete representation by rounding the output of the analysis transform, i.e., $\hatz = \lfloor f(\x) \rceil$, Yang et al. \cite{yang2020improving} cast the encoding problem as that of variational inference, and propose to infer the discrete representation that optimizes the per-data R-D cost. 
Their proposed method, SGA (Stochastic Gumbel Annealing), essentially solves a discrete optimization problem by constructing a categorical variational distribution $q(\z|\x)$ and optimizing w.r.t.\ its parameters by gradient descent, while annealing it to become deterministic so as to close the quantization gap \cite{yang2020improving}. In this work, we will adopt their proposed standalone procedure and opt to run SGA at test time, essentially treating it as a powerful black-box encoding procedure for a given NTC architecture.

\section{Methodology}
We begin with new empirical insight into the qualitative similarity between the synthesis transforms in NTC and traditional transform coding \cite{duan2022opening} (Sec.~\ref{sec:decoder-manifold-study}). 
This motivates us to adopt simpler synthesis transforms, such as JPEG-like block-wise linear transforms, which are computationally much more efficient than deep neural networks (Sec.~\ref{sec:decoder-design}). 
We then analyze the resulting effect on R-D performance and mitigate the performance drop using powerful encoding methods from the neural compression toolbox (Sec.~\ref{sec:nelbo-inference-gap-analysis}).

\subsection{The case for a shallow decoder} \label{sec:decoder-manifold-study}

Although the transforms in NTC are generally black-box deep CNNs, Duan et al. \cite{duan2022opening} showed that they in fact bear strong qualitative resemblance to the orthogonal transforms in traditional transform coding. They showed that the learned synthesis transform in various NTC architectures satisfy a certain separability property, i.e., a latent tensor can be decomposed spatially or across channels, then decoded separately, and finally combined in the pixel space to produce a reasonable reconstruction.
Moreover, decoding ``standard basis'' tensors in the latent space produces image patterns resembling the basis functions of orthogonal transforms.\footnote{We note that performing Principal Component Analysis on small image patches also results in similar patterns; see Figure~\ref{fig:filter-vis}  in the Appendix. }  

Here, we obtain new insights into the behavior of the learned synthesis transform in NTC. 
We show that the manifold of image reconstructions is approximately flat, in the sense that straight paths in the latent space are mapped to approximately straight paths (i.e., naive linear interpolations) in the pixel space. 
Additionally, the learned synthesis transform exhibits an approximate ``mixup'' 
 \cite{zhang2017mixup} behavior despite the lack of such explicit regularization during training.

\begin{figure}
    \centering
    \includegraphics[scale=1]{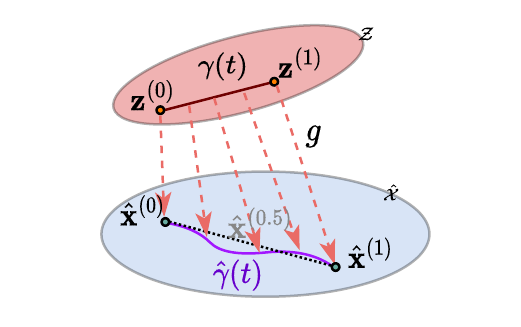}
    \caption{Conceptual illustration of the image manifold parameterized by $\hat \gamma(t)$ (purple curve), obtained by decoding a straight path $\gamma(t)$ in the latent space. We show it does not significantly deviate from a straight path (dashed line) connecting its two end points.}
    \label{fig:latent-traversal-conceptual}
\end{figure}

Suppose we are given an arbitrary pair of images $(\x^{(0)}, \x^{(1)})$, and we obtain
their latent coefficients $(\z^{(0)},  \z^{(1)})$ using the analysis transform (we ignore the effect of quantization as in Duan et al. \cite{duan2022opening}). Let $\gamma: [0, 1] \to \setZ$ be the straight path in the latent space defined by the two latent tensors, i.e., $\gamma(t) := (1-t) \z^{(0)} + t \z^{(1)}$. 
Using the synthesis transform $g$, we can then map the curve in the latent space to one in the space of reconstructed images, defined by $\hat \gamma(t) := g(\gamma(t))$. We denote the two end-points of the curve by $\hatx^{(0)} := g(\z^{(0)}) = \hat \gamma(0)$ and $\hatx^{(1)} := g(\z^{(1)}) = \hat \gamma(1)$. 
Instead of traversing the image manifold parameterized by $g$, we could also travel between the two end-points in a  straight path, which we define by $\hatx^{(t)} := (1-t) \hatx^{(0)} + t  \hatx^{(1)}$ and is given by a simple linear interpolation in the pixel space.  The setup is illustrated in Figure~\ref{fig:latent-traversal-conceptual}.

\begin{figure}[t]
\centering
\includegraphics[scale=0.42]{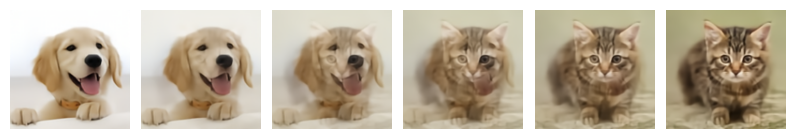}
\label{fig:xhat_ts}
\includegraphics[scale=0.42]{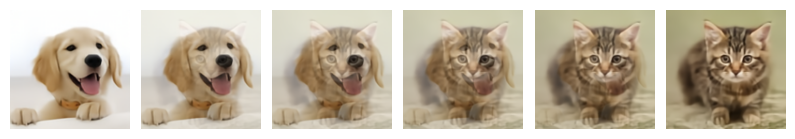}
\caption{Visualizing the 1-D manifold of image reconstructions $\{\hat\gamma(t) | t \in [0, 1]\}$ (\textbf{top row}) and the linear interpolation between its two end points, $\{(1-t) \hatx^{(0)} + t  \hatx^{(1)}| t \in [0, 1]\}$ (\textbf{bottom row}).
}
\label{fig:affine_xhat_ts}
\end{figure}

\begin{figure}[t]
\centering

     \begin{subfigure}[b]{0.28\textwidth}
         \centering
         \includegraphics[width=\linewidth]{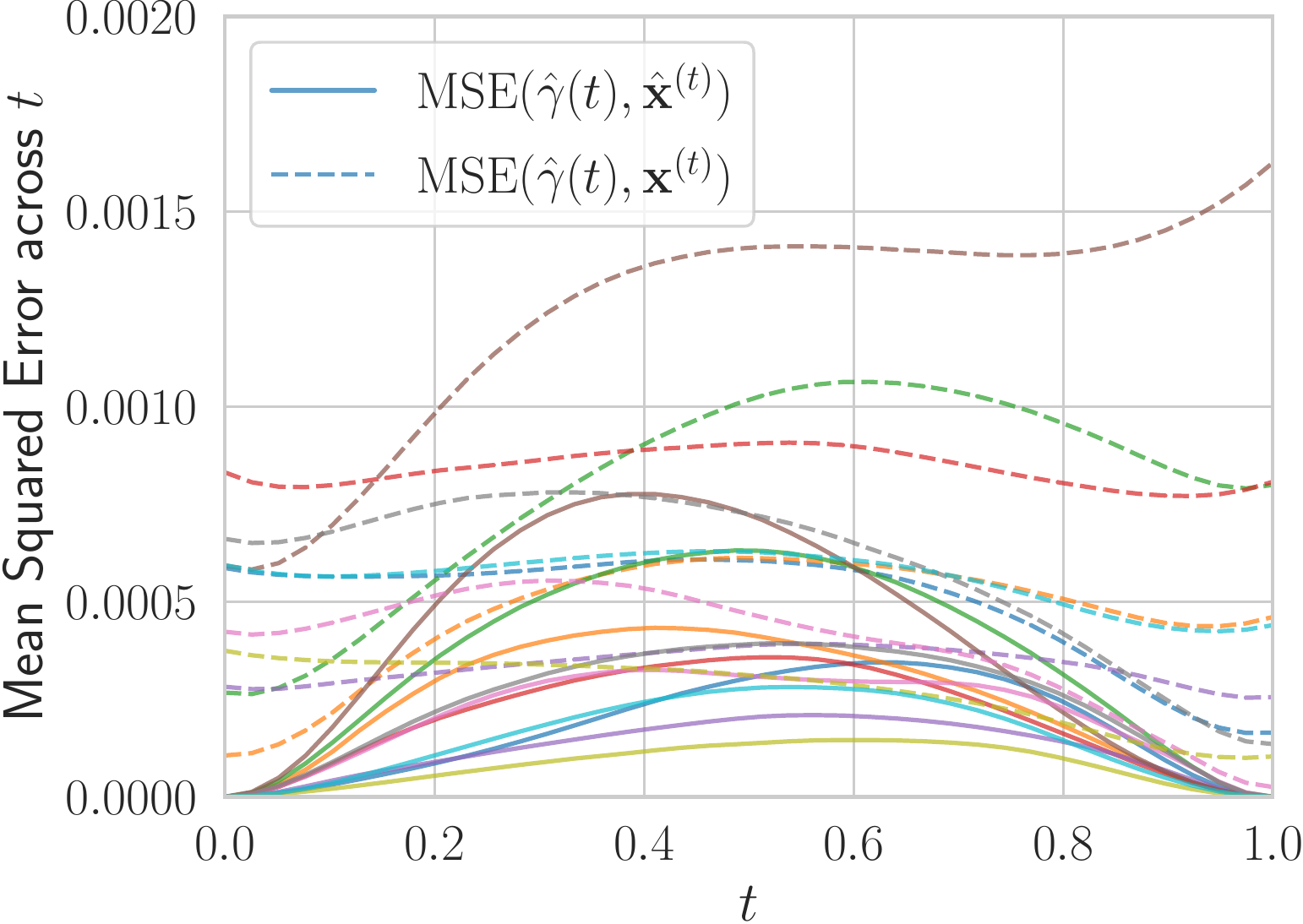}
         \caption{}\label{fig:latent_mses}
     \end{subfigure}
     \hfill
     \begin{subfigure}[b]{0.19\textwidth}
         \centering
         \includegraphics[width=\linewidth]{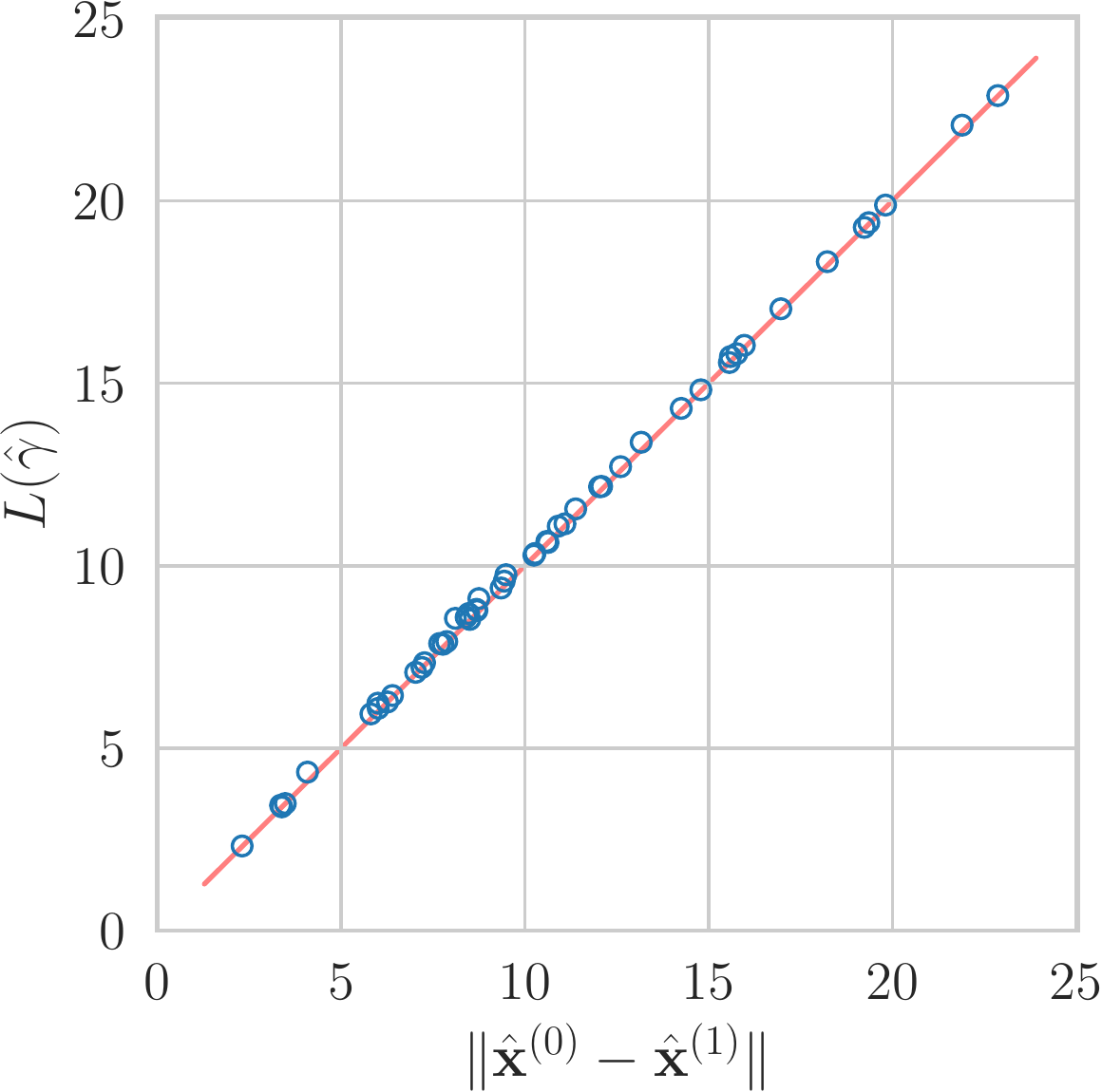}
         \caption{}\label{fig:curve_lengths_scatter}
     \end{subfigure}
\caption{
The effect of traversing the synthesis manifold, with end points defined by random image pairs.
\textbf{(a)}: Mean-squared error distance between the decoded curve $\hat\gamma(t)$ and straight paths in the image space (reconstructions $\hatx^{(t)}$ and originals $\x^{(t)}$). \textbf{(b)}: The length of the curve $\hat\gamma$ v.s. that of the interpolating straight path $\hatx^{(t)}$.
The image pixel values are scaled to $[-0.5, 0.5]$.
}
\label{fig:quantify-latent-traversal}
\end{figure}

Fig.~\ref{fig:affine_xhat_ts} visualizes an example of the resulting curve of images $\hat \gamma(t)$ (top row), compared to the interpolating straight path $\hatx^{(t)}$ (bottom row), as $t$ goes from 0 to 1. The results appear very similar, suggesting the latent coefficients largely carry local and mostly low-level information about the image signal.
As a rough measure of the deviation between the two trajectories,  Fig.~\ref{fig:latent_mses} computes the MSE between $\hat \gamma(t)$ and $\hatx^{(t)}$ at corresponding time steps, for pairs of random image crops from COCO \cite{lin2014microsoftcoco}. The results (solid lines) indicate that the two curves do not align perfectly. However, since the parameterization of any curve is not unique, we get a better sense of the behavior of the manifold curve $\hat\gamma(t)$ by considering its \emph{length} $L(\hat\gamma)$ in relation to the \emph{length} of the interpolating straight path $\|\hatx^{(0)} - \hatx^{(1)}\|$. We compute the two lengths (the curve length can be computed using the Jacobian of $g$; see Appendix Sec.~\ref{app:latent_traversal}), and plot them for  random image pairs in Fig.~\ref{fig:curve_lengths_scatter}. The resulting curve lengths fall very closely to the straight path lengths regardless of the absolute length of the curves, indicating that the curves globally follow nearly straight paths. 
Note that if $g$ was linear (affine), then $\hat\gamma(t)$ and $\hatx^{(t)}$ would  perfectly overlap  .

Additionally, inspired by \emph{mixup} regularization \cite{zhang2017mixup}, we examine how well the synthesized curve $\hat\gamma(t)$ can reconstruct the linear interpolation of the two \emph{ground truth} images, defined by $\x^{(t)} := ( 1- t) \x^{(0)} + t \x^{(1)}$.  Fig.~\ref{fig:latent_mses} plots the reconstruction error for the same random image pairs in dashed lines, and shows that
the synthesized curve $\hat\gamma(t)$ generally offers consistent reconstruction quality along the entire trajectory. Note that if $g$ was  linear (affine), then this reconstruction error would vary linearly across $t$.

The above observations form a stark contrast to the typical behavior of the decoder network in  generative modeling, where different images tend to be separated by regions of low density under the model, and the decoder function varies rapidly when crossing such boundaries \cite{chen2018metrics}, e.g., across a linear interpolation of images in pixel space.

We obtained these results with a Mean-scale Hyperprior model \cite{minnen2018joint} trained with $\lambda=0.01$, and we observe similar behavior at other bit-rates (with the curves $\hat\gamma$ becoming even ``straighter'' at higher bit-rates) and in various NTC architectures \cite{balle2017end, minnen2018joint, minnen2020channel} (see Appendix Sec.~\ref{app:additional_results} for more examples). 
Our empirical observations corroborate the earlier findings \cite{duan2022opening}, and raise the question: Given the many similarities, can we replace the deep convolutional synthesis in NTC with a linear (affine) function? 
Our motivation is mainly computational: a linear synthesis can offer drastic computation savings over deep neural networks. 
This is not necessarily the case for an arbitrary linear (affine) function from the latent to image space, so we restrict ourselves to efficient convolutional architectures.
As we show empirically in Sec.~\ref{sec:jpegl-experiment}, a single JPEG-like transform with a large enough kernel size can emulate a more general cascade of transposed convolutions, while being much more computationally efficient.
Compared to fixed and orthogonal transforms in traditional transform coding, learning a linear synthesis from data allows us to still benefit from end-to-end optimization. 
Further, in Sec.~\ref{sec:main-experiment-results}, we show that strategically incorporating a small amount of nonlinearity can significantly improve the R-D performance without much increase in computation complexity.

\subsection{Shallow decoder design}\label{sec:decoder-design}

\paragraph{JPEG-like synthesis}
At its core, JPEG works by dividing an input image into $8 \times 8$ blocks and applying block-wise linear transform coding. This can be implemented efficiently in hardware and is a key factor in JPEG's enduring popularity. 
By analogy to JPEG, we interpret the $h\times w \times C$ latent tensor in NTC as the coefficients of a linear synthesis transform. In the most basic form, the output reconstructions are computed in $s \times s$ blocks, similarly to JPEG. %
Specifically, the $(i, j)$th block reconstruction is computed as a linear combination of (learned) ``basis images'' $\mathbf{K}_c \in \mathbb{R}^{s \times s \times C_{out}}, c=1, ..., C$, weighted by the vector of  (quantized) coefficients $\mathbf{z}_{i,j} \in \mathbb{R}^C$ associated with the $(i, j)$th spatial location:
\begin{align}
    \hat{B}_{i, j} = \sum_{c=1}^C \mathbf{z}_{i,j,c} \mathbf{K}_c. \label{eq:jpeg-syn}
\end{align}

Note that we recover the per-channel discrete cosine transform of JPEG by setting $s=8, C=64, C_{out}=1$, and $\{ \mathbf{K}_c, c=1, ..., 64 \}$ to be the bases of the $8\times 8$ discrete cosine transform. \quad
Eq~\ref{eq:jpeg-syn} can be implemented efficiently via a transposed convolution on $\mathbf{z}$, using $\mathbf{K}$ as the kernel weights and $s$ as the stride. 
In terms of MACs, the computation complexity of the JPEG-like synthesis then equals
\begin{align}
    M(\text{JPEG-like}) =  C \times h \times w \times s^2  \times C_{out} , \label{eq:jpegl-macs-formulae}
\end{align}
where $C_{out} = 3$ for a color image.\footnote{When the latent coefficients are sparse (which often occurs at low bit-rates), this computation complexity can be further reduced by using sparse matrix/tensor operations. We leave this to future work. }
Note that for a given latent tensor and ``upsampling'' rate $s$, Eq.~\ref{eq:jpegl-macs-formulae} gives the \emph{minimum achievable} MACs by any non-degenerate synthesis transform based on (transposed) convolutions.
As we see in Sec.~\ref{sec:main-experiment-results}, 
although the minimal JPEG-like synthesis drastically reduces the decoding complexity, it can introduce severe blocking artifacts since the blocks are reconstructed independently. 
We therefore allow overlapping basis functions with spatial extent $k \times k$, where $k \geq s$ and $k-s$ is the number of overlapping pixels; we compute each $k \times k$ blocks as in Eq.~\ref{eq:jpeg-syn}, then form the reconstructed image by taking the sum of the (overlapping) blocks. This corresponds to simply increasing the kernel size from $(s,s)$ to $(k,k)$ in the corresponding transposed convolution, and increases the $s^2$ factor in Eq.~\ref{eq:jpegl-macs-formulae} to $k^2$.

\paragraph{Two-layer nonlinear synthesis}
Despite its computational efficiency, the JPEG-like synthesis can be overly restrictive.
Indeed, nonlinear transform coding benefits from the ability of the synthesis transform to adapt to the shape of the data manifold \cite{balle2021ntc}. 
We therefore introduce a small degree of nonlinearity in the JPEG-like transform. Many possibilities exist, and we found that
introducing a single hidden layer with nonlinearity to work well. 
Concretely, we use two layers of transposed convolutions $(\texttt{conv\_1}, \texttt{conv\_2})$, with strides $(s_1, s_2)$, kernel sizes $(k_1, k_2)$, and output channels $(N, C_{out})$ respectively.  
At lower bit-rates, we found it more parameter- and compute-efficient to also allow a residual connection from $\z$ to the hidden activation using another transposed convolution $\texttt{conv\_res}$ (see a diagram and more details in Appendix Sec.~\ref{app:2layer-res-arch-details}). Thus, given a latent tensor $\z \in \mathbb{R}^{h, w, C}$
the output is $g(\z) = \texttt{conv\_2} ( \xi (\texttt{conv\_1}(\z)) + \texttt{conv\_res}(\z))$, where $\xi$ is a nonlinear activation.

The MAC count in this architecture is then approximately
\begin{align}
    M(\text{2-layer}) &=  C \times h \times w \times k_1^2 \times 2N \\ \nonumber
    &+ N \times h s_1\times w s_1 \times k_2^2 \times C_{out}.
\end{align}
To keep this decoding complexity low, we use large convolution kernels ($k_1=13$) with aggressive upsampling ($s_1=8$) in the first layer, in the spirit of a JPEG-like synthesis, followed by a lightweight output layer with a smaller upsampling factor ($s_2=2$) and kernel size ($k_2=5$).  
We use the simplified (inverse) GDN activation \cite{johnston2019computationally} for $\xi$ as it gave the best R-D performance with minor computational overhead. We discuss these and other architectural choices in Sec.~\ref{sec:ablation-experiments}.

\subsection{Formalizing the role of the encoder in lossy compression performance}\label{sec:nelbo-inference-gap-analysis}
Here, we analyze the rate-distortion performance of neural lossy compression in an idealized, asymptotic setting. Our novel decomposition of the R-D objective pinpoints the performance loss caused by restricting to a simpler (e.g., linear) decoding transform, and suggests reducing the inference gap as a simple and theoretically principled remedy. 

Consider a general neural lossy compressor operating as follows. Let $\setZ$ be a latent space, $p(\z)$ a prior distribution over $\setZ$ known to both the sender and receiver, and
$g: \setZ \to \sethX$ the synthesis transform belonging to a family of functions $\mathcal{G}$. 
Given a data point $\x$, the sender computes an inference distribution $q(\z | \x)$; this can be the output of an encoder network, or a more sophisticated procedure such as iterative optimization with SGA \cite{yang2020improving}. 
We assume relative entropy coding \cite{cuff2008, theis2021algorithms} is applied with minimal overhead, %
so that the sender can send a sample of $\z \sim q(\z | \x)$ with an average bit-rate not much higher than $KL(q(\z | \x) \| p(\z))$ \cite{flamich2020cwq}. 
Given a neural compression method, which can be identified with the tuple  $(q(\z|\x), g, p(\z))$, %
its R-D cost on data distributed according to $P_\X$ thus has the form of a negative ELBO \cite{flamich2020cwq}
\begin{align}
    &\mathcal{L}(q(\z | \x), g, p(\z)) :=\\ \nonumber
    & \lambda \E_{\x \sim P_\X, \z \sim q(\z|\x)} [\rho(\x, g(\z))] + \E_{\x \sim P_\X}[ KL(q(\z | \x) \| p(\z)) ],
\end{align}
where $\lambda \geq 0$ controls the R-D tradeoff, and $\rho: \setX \times \sethX \to [0, \infty)$ is the distortion function (commonly the MSE).
Note that the encoding distribution $q(\z | \x)$ appears in both the rate and distortion terms above. 
We show that the compression cost admits the following alternative decomposition, where the effects of $p(\z)$, $g$, and $q(\z | \x)$ can be isolated:
\begin{align}
    &\mathcal{L}(q(\z | \x), g, p(\z)) :=\\ \nonumber
    & = \underbrace{\mathcal{F}(\mathcal{G})}_{\text{irreducible}} + \underbrace{\left(\E_{\x \sim P_\X}[- \log \Gamma_{g, p(\z)} (\x)] - \mathcal{F}(\mathcal{G})\right)}_{\text{modeling gap}} \\ 
    & + \underbrace{\E_{\x\sim P_\X}[KL(q(\z|\x) \| p(\z|\x)]}_{\text{inference gap}}. \label{eq:rd-decomposition}
\end{align}
The derivation and definition of various quantities are given in Sec.~\ref{app:rd-cost-decomposition}, and mirror a similar decomposition in lossless compression \cite{zhang2022generalization}; here we give a high-level explanation of the three terms.  The first term represents the fundamentally irreducible cost of compression; this depends only on the intrinsic compressibility of the data $P_\X$ \cite{yang2022towards} and the transform family $\mathcal{G}$. The second term represents the excess cost of compression given our particular choice of decoding architecture, i.e., the prior $p(\z)$ and transform $g$, compared to the optimum achievable (the first term); we thus call it the \emph{modeling gap}. Note that for each choice of $(g, p(\z))$, the optimal inference distribution has an explicit formula, which allows us to write the R-D cost under optimal inference in the form of a negative log partition function (the $-\log \Gamma$ term).  Finally, we consider the effect of suboptimal inference and isolate it in the third term,  representing the overhead caused by a sub-optimal encoding/inference method $q(\z | \x)$ for a given model $(g, p(\z))$; we call it the \emph{inference gap}.  

Although the above result is derived in an asymptotic setting, it still gives us insight about the performance of neural lossy compression at varying decoder complexity. 
When we use a simpler synthesis transform architecture, we place restrictions on our transform family $\mathcal{G}$, thus causing the first (irreducible) part of compression cost to increase. %
The modeling gap may or may not increase as a result,\footnote{The modeling gap can be reduced by adopting a more expressive prior $p(\z)$, although doing so can lead to higher decoding complexity. } but we can always lower the overall compression cost by reducing the inference gap, without affecting the decoding computational complexity.

In this work, we explore two orthogonal approaches for reducing the inference gap, which can be further decomposed into an (1) \emph{approximation gap} and (2) \emph{amortization gap} \cite{cremer2018inference}. 
Correspondingly, for a given decoding architecture, we propose to reduce (1) by using a more powerful analysis transform, e.g., from a recent SOTA method such as ELIC \cite{he2022elic}, and reduce (2) by performing iterative encoding using SGA \cite{yang2020improving} at compression time.

\section{Experiments}
\subsection{Data and training}
We train all of our models on random $256 \times 256$ image crops from the COCO 2017 \cite{lin2014microsoftcoco} dataset. We follow the standard training procedures as in \cite{balle2017end, minnen2018joint} and optimize for MSE as the distortion metric.
We verified that our base Mean-Scale Hyperprior model matches the reported performance in the original paper \cite{minnen2018joint}.

\begin{table*}[ht]
\begin{tabular}{ccccccccc}
\toprule
\multirow{2}{*}{Method} & 
     \multicolumn{6}{c}{Computational complexity (KMAC)} & \multirowcell{2}{Syn. param\\ count (Mil.)} & \multirowcell{2}{BD rate\\ savings (\%) ↑} \\ \cline{2-7}
 & $f$ &  $f_h$ &  enc. tot. &    $g$ &  $g_h$ & dec. tot. \\
\midrule
He 2022 ELIC \cite{he2022elic}                 & 255.42 &   6.73 &     262.15 & 255.42 & 126.57 &     381.99 &               7.34 &            26.98 \\
Minnen 2020 CHARM \cite{minnen2020channel}            &  93.79 &   5.90 &      99.70 &  93.79 & 256.51 &     350.30 &               4.18 &            20.02 \\
Wang 2023 EVC \cite{wang2023EVC}                & 263.25 &   1.86 &     265.11 & 257.94 &  34.82 &     292.76 &               3.38 &            22.56 \\
Minnen 2018 Hyperprior \cite{minnen2018joint}       &  93.79 &   6.73 &     100.52 &  93.79 &  15.18 &     108.97 &               3.43 &             3.30 \\
Ballé 2017 Factorized Prior \cite{balle2017end}  &  81.63 &   0.00 &      81.63 &  81.63 &   0.00 &      81.63 &               3.39 &           -32.93 \\
2-layer syn. + SGA (proposed) & 255.42 &   6.73 &     \textasciitilde10\textsuperscript{5} &   \textbf{5.34} &  15.18 &      20.52 &               \textbf{1.30} &             4.67 \\
2-layer syn. (proposed)       & 255.42 &   6.73 &     262.15 &   \textbf{5.34} &  15.18 &      20.52 &               \textbf{1.30} &            -5.19 \\
JPEG-like syn. (proposed)     & 255.42 &   6.73 &     262.15 &   \textbf{1.22} &  15.18 &      16.39 &               \textbf{0.31} &           -20.95 \\
\bottomrule
\end{tabular}
\caption{Computational complexity of various neural compression methods, v.s. average BD rate savings relative to BPG \cite{bellard2014bpg} on Kodak.
Complexity is measured in KMACs (thousand multiply–accumulate operations) per pixel, and does not include entropy coding.
$f, f_h, g, g_h$ stand for analysis, hyper analysis, synthesis, and hyper synthesis transforms. We also report the parameter count of synthesis transforms ($g$) in the second-to-last column, and a rough estimate of the overall encoding complexity of SGA-based encoding (\textasciitilde10\textsuperscript{5} KMACs/pixel).
\label{tab:macs}
}
\end{table*}

\begin{figure*}[ht]
     \centering
     \begin{subfigure}[b]{0.45\textwidth}
         \centering
         \includegraphics[width=\textwidth]{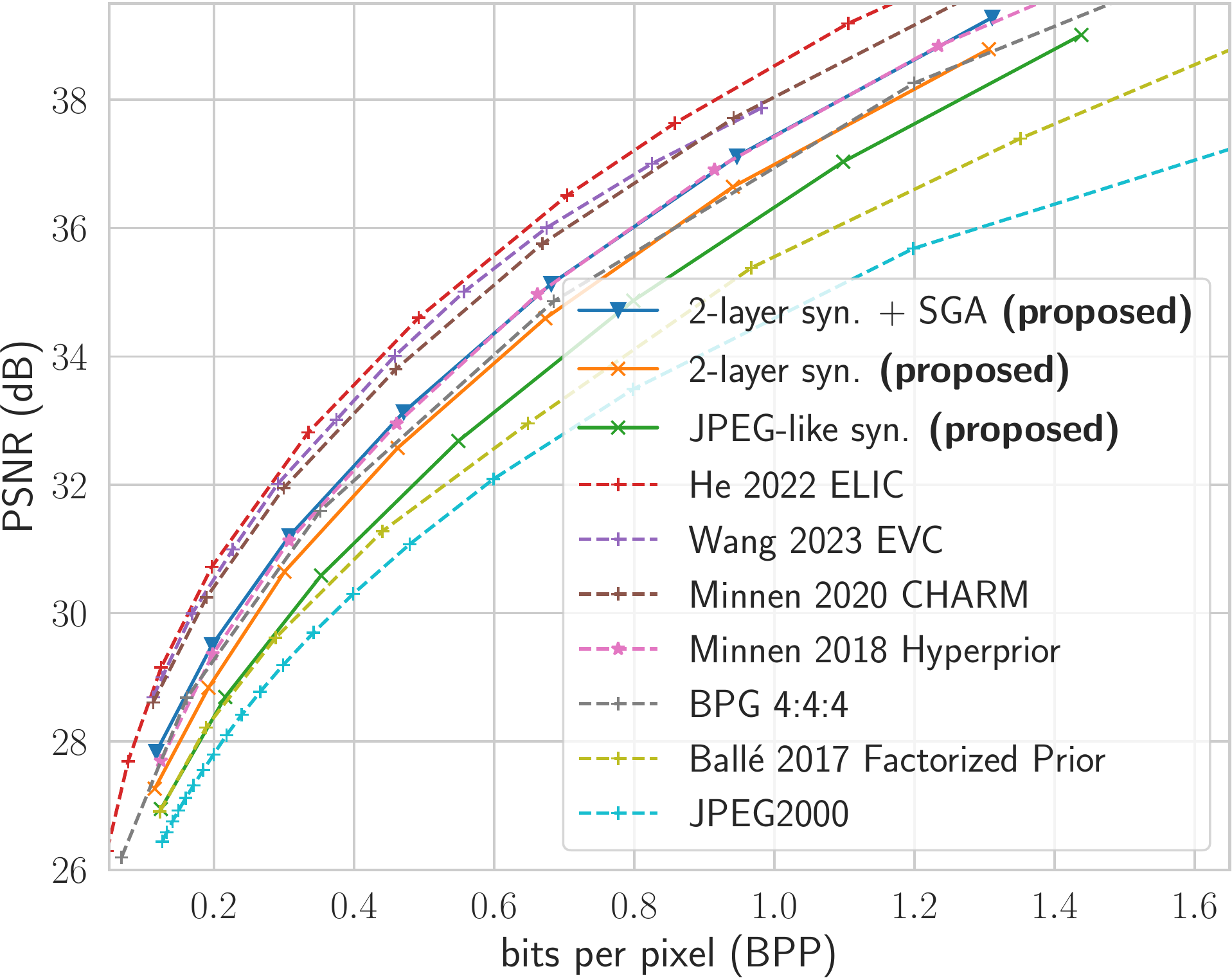}
         \caption{R-D performance on Kodak; quality measured in PSNR (the higher the better). }
         \label{fig:rd_psnr-kodak}
     \end{subfigure}
     \hfill
     \begin{subfigure}[b]{0.45\textwidth}
         \centering
         \includegraphics[width=\textwidth]{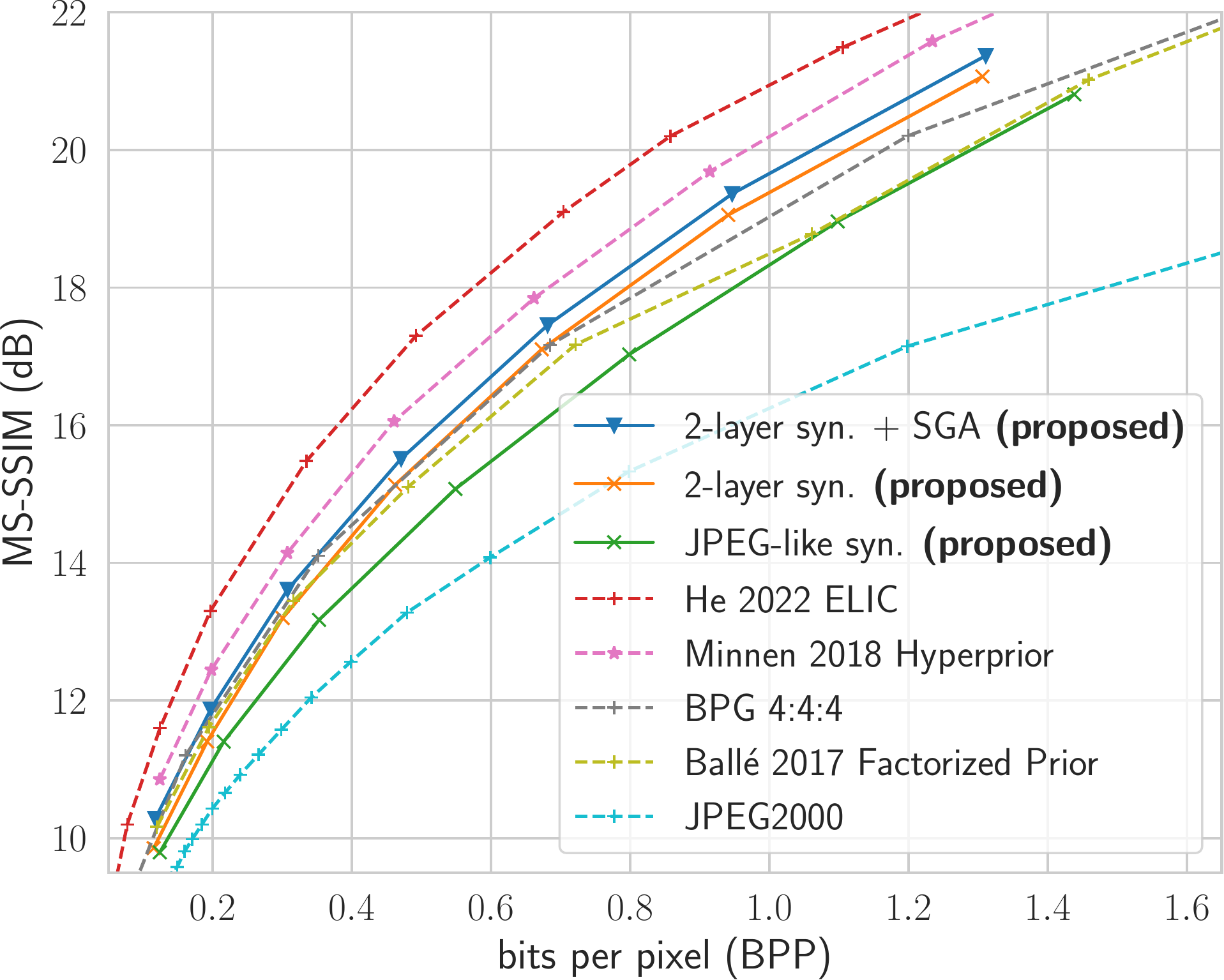}
         \caption{R-D performance on Kodak; quality measured in MS-SSIM dB (the higher the better). }
         \label{fig:rd_msssim-kodak}
     \end{subfigure}
        \caption{Comparison of the R-D performance of the proposed methods with existing neural image compression methods. All the models were optimized for MSE distortion.}
        \label{fig:rd_kodak}
\end{figure*}

\subsection{Comparison with existing methods} \label{sec:main-experiment-results}

We compare our proposed methods with standard neural compression methods \cite{balle2017end, minnen2018joint, minnen2020channel} and state-of-the-art methods \cite{he2022elic, wang2023EVC} targeting computational efficiency. We obtain the baseline results from the CompressAI library \cite{begaint2020compressai}, or trace the results from papers when they are not available.  For our shallow synthesis transforms, we use $k=18$ in the JPEG-like synthesis, and $N=12, k_1=13, k_2=5$ in the 2-layer synthesis; we ablate on these choices in Sec.~\ref{sec:jpegl-experiment} and \ref{sec:ablation-experiments}.

Table~\ref{tab:macs} summarizes the computational complexity of various methods, ordered by decreasing overall decoding complexity. We use the \texttt{keras-flops} package \footnote{\url{https://pypi.org/project/keras-flops/}} to measure the FLOPs on $512 \times 768$ images, and report the results in KMACs (thousand multiply-accumulates) per pixel. 
Note that the Factorized Prior architecture \cite{balle2017end} lacks the hyperprior, 
while CHARM \cite{minnen2020channel} and ELIC \cite{he2022elic} use autoregressive computation in the hyperior. Our proposed models borrow the same hyperprior from Mean-Scale Hyperprior \cite{minnen2018joint}.

While most existing methods use analysis and synthesis transforms with symmetric computational complexity, our proposed methods adopt the relatively more expensive analysis transform from ELIC \cite{he2022elic} (column ``$f$''), and drastically reduces the complexity of the synthesis transform (column ``$g$'')
 --- over 50 times smaller than in ELIC, and 17 smaller than in Mean-Scale Hyperprior.\footnote{In our preliminary measurements, this translates to 6 \textasciitilde 12 times reduction in running time of the synthesis transform compared to the Mean-Scale Hyperprior, depending the hardware used.} As a result, the hyper synthesis transform (the same as in Mean-Scale Hyperprior) accounts for a great majority of our overall decoding complexity.

\begin{figure*}[h]
     \centering
     \begin{subfigure}[b]{0.24\textwidth}
         \centering
         \includegraphics[width=\textwidth]{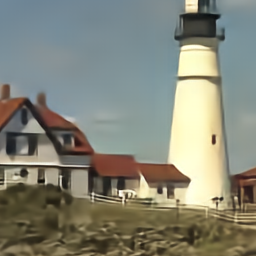}
         \label{fig:y equals x}
     \end{subfigure}
     \hfill
     \begin{subfigure}[b]{0.24\textwidth}
         \centering
         \includegraphics[width=\textwidth]{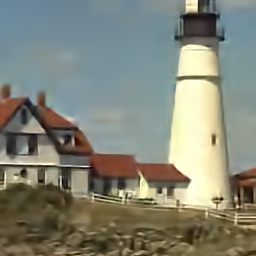}
         \label{fig:three sin x}
     \end{subfigure}
     \hfill
     \begin{subfigure}[b]
     {0.24\textwidth}
         \centering
         \includegraphics[width=\textwidth]{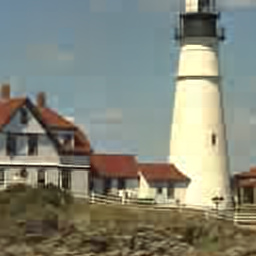}
         \label{fig:three sin x}
     \end{subfigure}
     \hfill
     \begin{subfigure}[b]{0.24\textwidth}
         \centering
         \includegraphics[width=\textwidth]{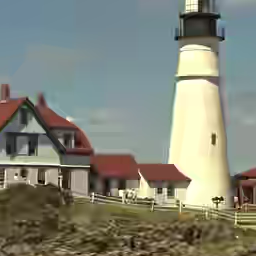}
         \label{fig:five over x}
     \end{subfigure}
        \caption{Visualizing the different kinds of distortion artifacts at comparable low bit-rates between various methods. Left to right: Mean-Scale Hyperprior \cite{minnen2018joint}, two-layer synthesis (proposed), JPEG-like synthesis (proposed), and BPG \cite{bellard2014bpg}. See Sec.~\ref{sec:main-experiment-results} for relevant discussion. }
        \label{fig:main_qualitative}
\end{figure*}

In Fig.~\ref{fig:rd_psnr-kodak}, we plot the R-D performance of various methods on the Kodak \cite{kodak} benchmark, with quality measured in PSNR. 
We also compute the BD \cite{bjontegaard2001calculation} rate savings (\%) relative to BPG \cite{bellard2014bpg}, and summarize the average BD rate savings v.s.\ the total decoding complexity in Table~\ref{tab:macs} and Fig.~\ref{fig:rdc_kodak}.
As can be seen, our model with ELIC analysis transform and JPEG-like synthesis transform (\textbf{green}) 
 comfortably outperforms the Factorized Prior architecture \cite{balle2017end}; the latter employs a more expensive CNN synthesis transform but a less powerful entropy model. However, our JPEG-like synthesis still significantly lags behind BPG and the Mean-Scale Hyperprior. By adopting the two-layer synthesis (\textbf{orange}), the overall decoding complexity increases marginally (since the majority of complexity comes from the hyper decoder), while the R-D performance improves significantly, to within $\leq 6\%$ bit-rate of BPG.  Finally, performing iterative encoding with SGA  (\textbf{blue}) gives a further boost in R-D performance, outperforming the Mean-Scale Hyperprior (and BPG) without incurring any additional decoding complexity.

Additionally, we examine the R-D performance using the more perceptually relevant MS-SSIM metric \cite{wang2003msssim}. Following standard practice,  we display it in dB as $-10 \log_{10} ( 1 - \text{MS-SSIM})$. The results are shown in Fig.~\ref{fig:rd_msssim-kodak}. We observe largely the same phenomenon as before under PSNR, except that the existing methods based on CNN decoders achieve relatively much stronger performance compared to traditional codecs such as BPG and JPEG 2000. Our proposed method with a two-layer synthesis and iterative encoding (\textbf{blue}) still outperforms BPG, but no longer outperforms the Mean-Scale Hyperprior (\textbf{pink}).
Indeed, as we see in Fig.~\ref{fig:main_qualitative}, the reconstructions of the proposed shallow synthesis transforms can exhibit artifacts similar to classical codecs (e.g., BPG) at low bit-rates, such as blocking or ringing, but to a lesser degree with the nonlinear two-layer synthesis (second panel) than  the JPEG-like synthesis (third panel).

In Sec.~\ref{sec:additional-rd} of the Appendix, we report additional R-D results evaluated on Tecnick \cite{tecnick} and the CLIC validation set \cite{clic2018}, as well as under the perceptual distortion LPIPS \cite{zhang2018lpips}. Overall, we find that our proposed two-layer synthesis with SGA encoding matches the Hyperprior performance when evaluated on PSNR, but under-performs by $8\% \sim 12 \%$ (in BD-rate) when evaluated on either MS-SSIM or LPIPS. 

\subsection{JPEG-like synthesis} \label{sec:jpegl-experiment}
In this section, we study the JPEG-like synthesis in isolation. We start with the Mean-Scale Hyperprior architecture, and replace its CNN synthesis with a single transposed convolution with varying kernel sizes. Additionally, instead of replacing the CNN synthesis entirely, we also consider a linear version of it (``linear CNN synthesis'') where we remove all the nonlinear activation functions. This results in a composition of four transposed convolution layers, which in general cannot be expressed by a single transposed convolution;  however, note that this is still a linear (afffine) map from the latent space to image space.

\begin{figure*}[ht]
     \centering
     \begin{subfigure}[b]{0.24\textwidth}
         \centering
         \includegraphics[width=\textwidth]{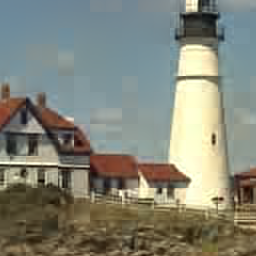}
         \label{fig:y equals x}
     \end{subfigure}
     \hfill
     \begin{subfigure}[b]{0.24\textwidth}
         \centering
         \includegraphics[width=\textwidth]{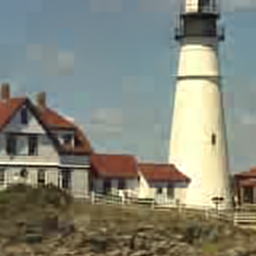}
         \label{fig:three sin x}
     \end{subfigure}
     \hfill
     \begin{subfigure}[b]
     {0.24\textwidth}
         \centering
         \includegraphics[width=\textwidth]{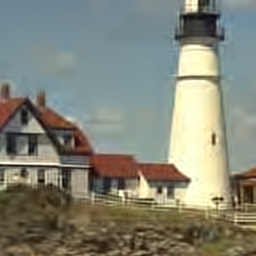}
         \label{fig:three sin x}
     \end{subfigure}
     \hfill
     \begin{subfigure}[b]{0.24\textwidth}
         \centering
         \includegraphics[width=\textwidth]{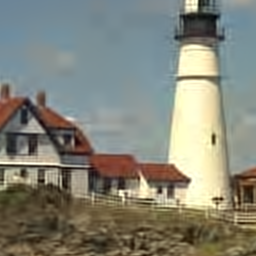}
         \label{fig:five over x}
     \end{subfigure}
        \caption{Comparing the distortion artifacts at low bit-rate for different kernel sizes ($k=16, 18, 32$, from left to right) in our JPEG-like synthesis, as well as a linear CNN synthesis (rightmost panel). The JPEG-like blocking artifacts are reduced as $k$ increases; see Sec.~\ref{sec:jpegl-experiment}. }
        \label{fig:jpegl_ablation_qualitative}
\end{figure*}

Fig.~\ref{fig:jpegl_ablation_qualitative} illustrates the distortion artifacts of the JPEG-like synthesis and linear CNN synthesis at comparable bit-rates, and reveals the following: (i). Using the smallest non-degenerate kernel size ($k=s=16$) results in severe blocking artifacts, e.g., as seen in the $16 \times 16$ cloud patches in the sky, similarly to JPEG. (ii). Increasing $k$ by a small amount (16 → 18) already helps smooth out the blocking, but further increase gives diminishing returns. (iii). At $k=32$, the reconstruction of JPEG-like synthesis no longer shows obvious blocking artifacts, but shows ringing artifacts near object boundaries instead; the reconstruction by the linear CNN synthesis gives visually very similar results.   %

\begin{figure}[h]
    \centering
    \includegraphics[width=0.7\linewidth]{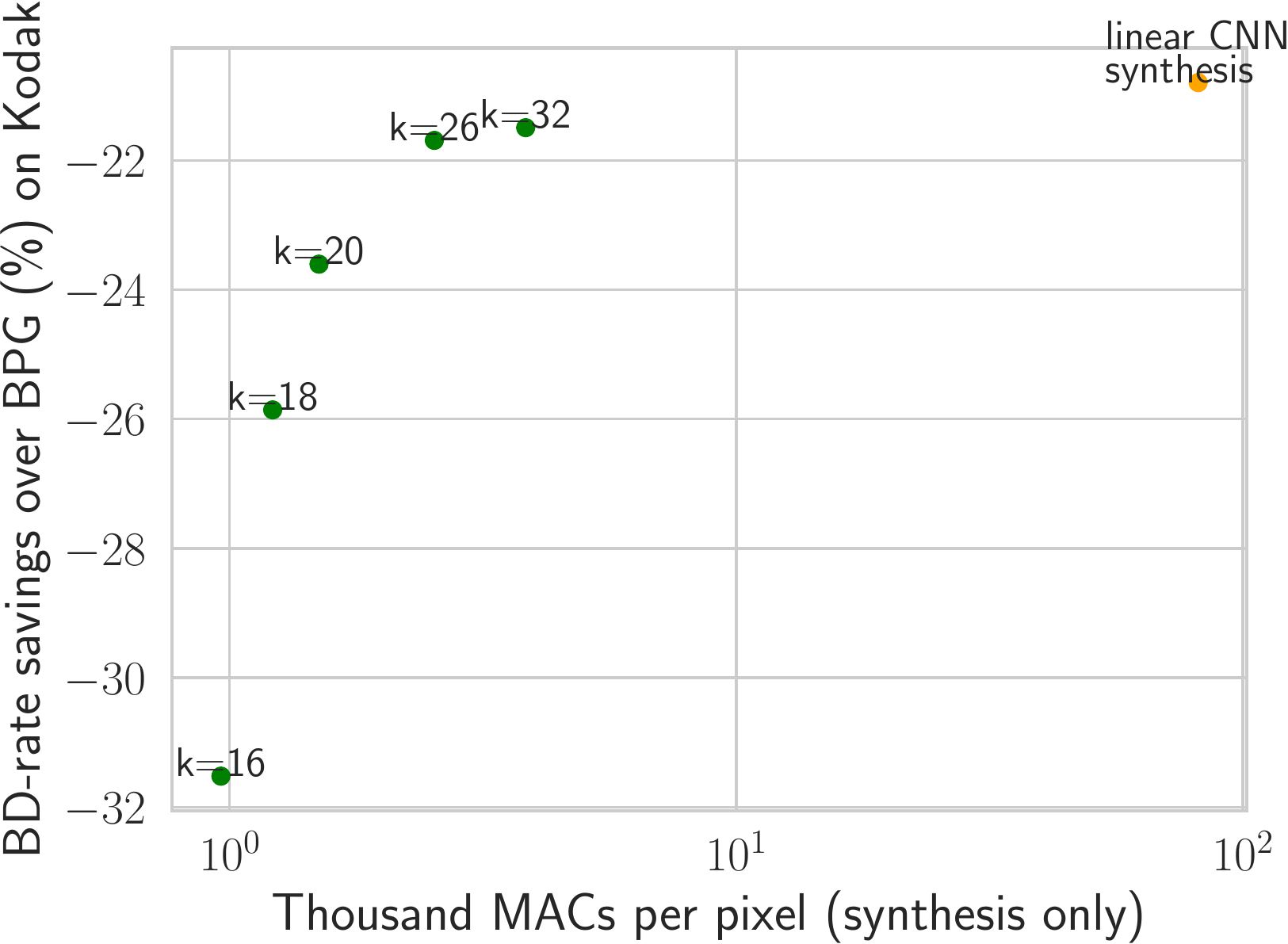}
    \caption{Effect of increasing kernel size ($k$) on the performance of JPEG-like synthesis. See Sec.~\ref{sec:jpegl-experiment} for details.}
    \label{fig:jpegl_ablation}
\end{figure}

Indeed, Fig.~\ref{fig:jpegl_ablation} confirms that increasing $k$ quantitatively improves the R-D performance of the JPEG-like synthesis, with $k = 26$ approaching the R-D performance of the linear CNN synthesis (within 1\% aggregate bit-rate) while requiring 94\% less FLOPs. 
We conclude that for image compression, a single transposed convolution with large enough kernel size can largely emulate a deep but linear CNN in PSNR performance, and the additional nonlinearity is necessary for  the superior perceptual quality of nonlinear transform coding.

\subsection{Ablation studies}\label{sec:ablation-experiments}

\paragraph{The analysis transform.}
We ablate on the choice of analysis transform for our proposed two-layer synthesis architecture. 
Replacing the analysis transform of ELIC \cite{he2022elic} with that of  Mean-Scale Hyperprior results in over 6\% worse bitrate (with BPG as the anchor). This gap can be reduced to \textasciitilde 5\% by increasing the number of base channels in the CNN analysis, although with diminishing returns and becomes suboptimal compared to switching to the ELIC analysis transform.
See Appendix Sec.~\ref{app:more-ablations} for details.

\paragraph{Two-layer synthesis architecture}
Due to resource constraints, we were not able to conduct an exhaustive architecture search, and instead set the hyperparameters manually.

\begin{figure}[ht]
     \centering
     \begin{subfigure}[b]{0.23\textwidth}
         \centering
         \includegraphics[width=\linewidth]{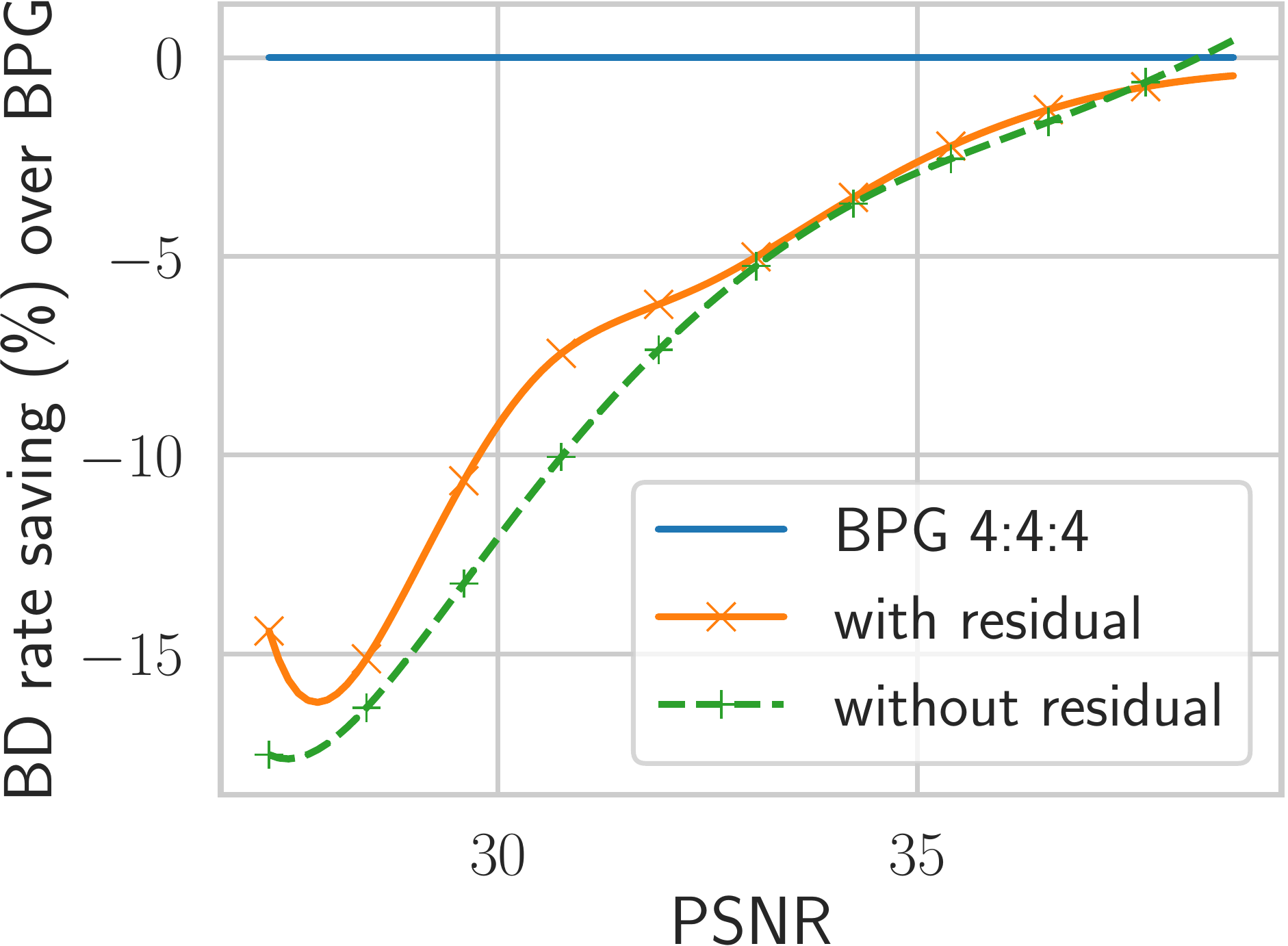}
         \caption{Residual connection.}
     \end{subfigure}
     \hfill
     \begin{subfigure}[b]{0.23\textwidth}
         \centering
         \includegraphics[width=\linewidth]{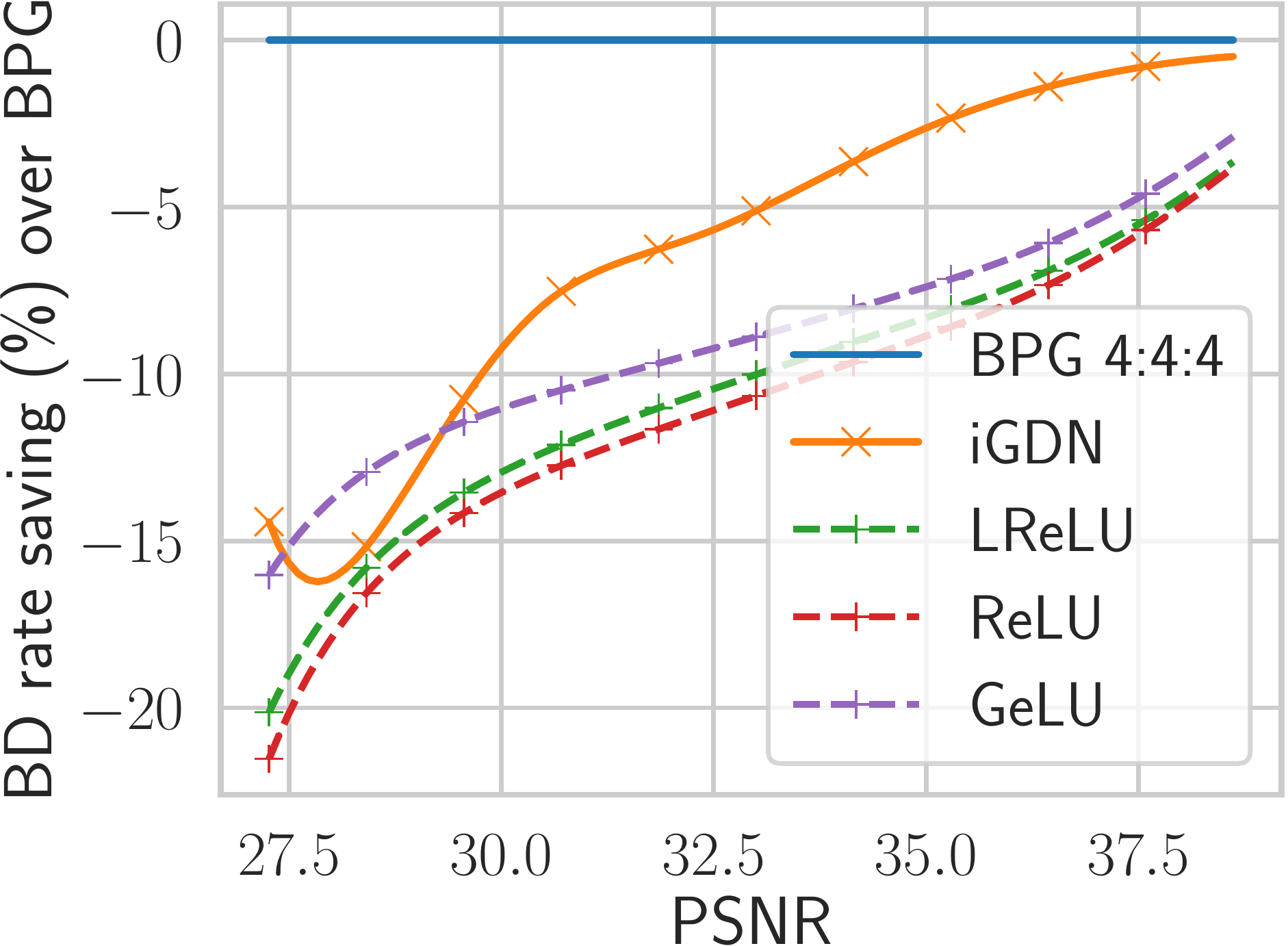}
         \caption{Nonlinear activation. }
     \end{subfigure}
     \hfill
     \begin{subfigure}[b]{0.23\textwidth}
         \centering
        \includegraphics[width=\linewidth]{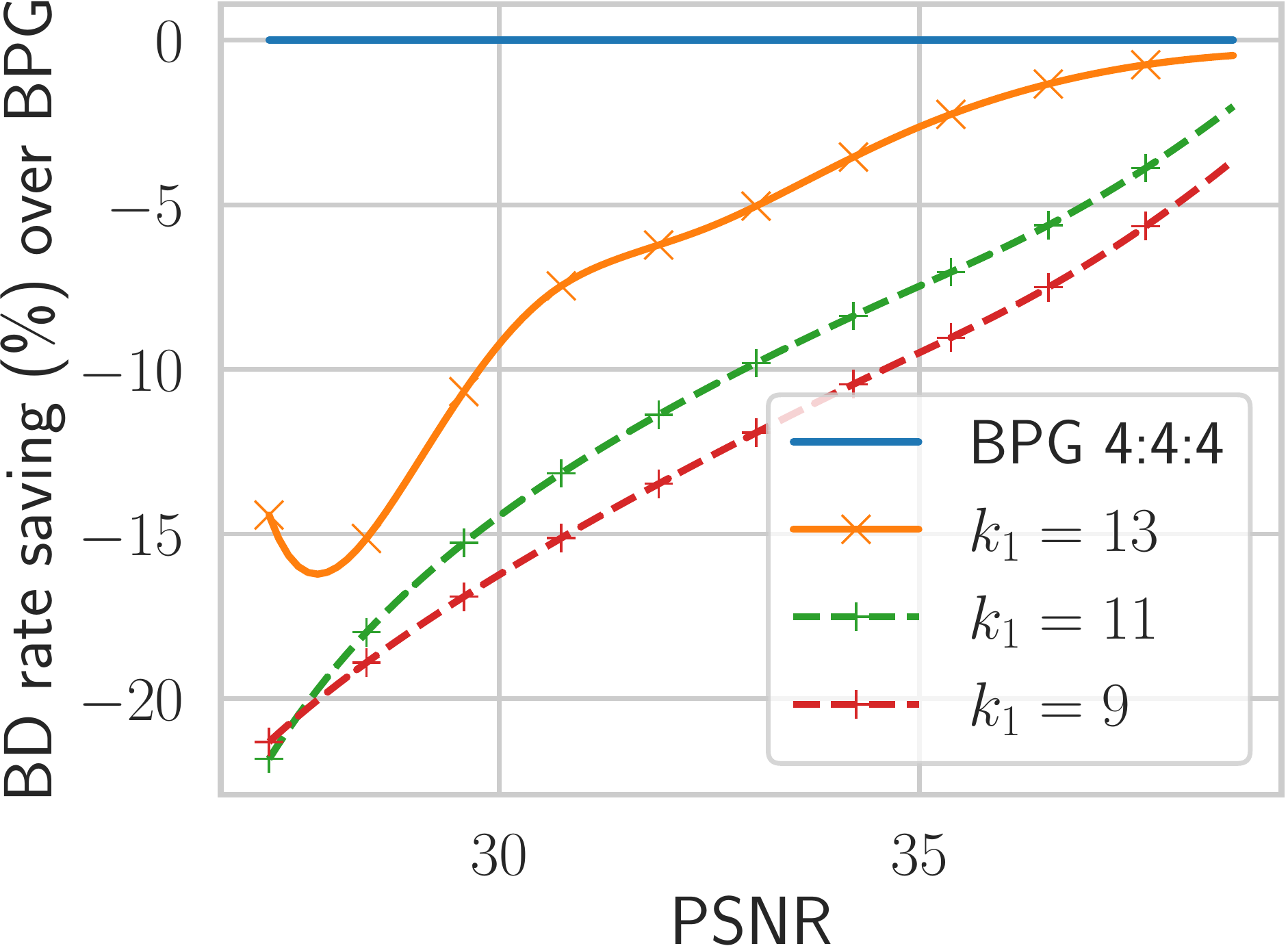}
         \caption{Hidden layer kernel size.}
     \end{subfigure}
     \hfill
     \begin{subfigure}[b]{0.23\textwidth}
         \centering
   \includegraphics[width=\linewidth]{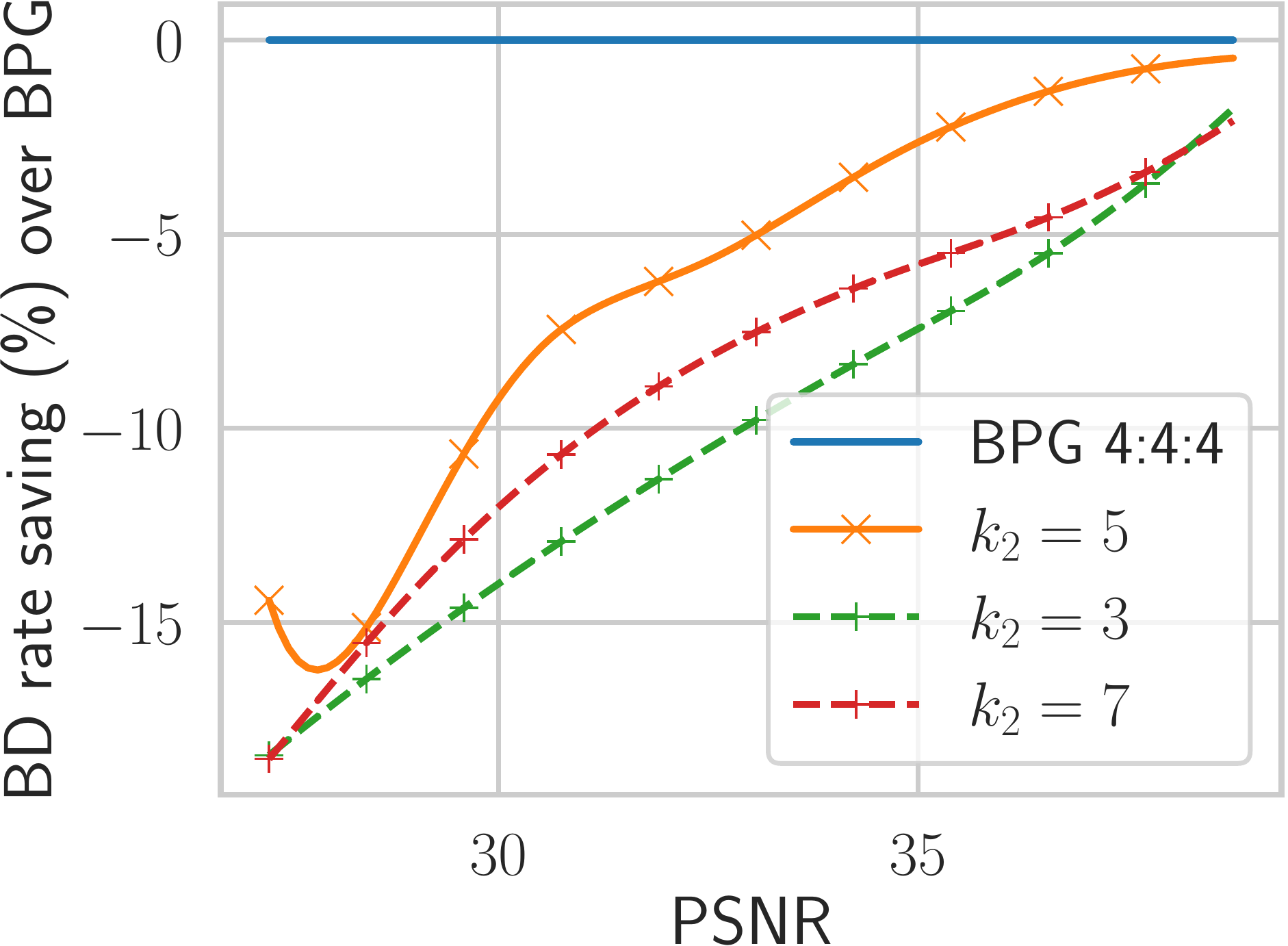}
         \caption{Output layer kernel size. }
     \end{subfigure}
        \caption{Ablation on various architectural choices of the proposed two-layer synthesis transform. 
        BD-rate savings are evaluated on Kodak (the higher the better).
        See Sec.~\ref{sec:ablation-experiments} for a discussion.}
        \label{fig:ablation_shallow_arch}
\end{figure}

Fig.~\ref{fig:ablation_shallow_arch} presents ablation results on the main architectural elements of the proposed two-layer synthesis. We found that the residual connection slightly improves the R-D performance at low bit-rates, compared to a simple two-layer architecture with comparable FLOPs (using $2N=24$ hidden channels).
We also found the use of (inverse) GDN activation \cite{balle2016end} and increased kernel size in the output layer ($k_2$) to be beneficial, which only cost a minor (less than 5\%) increase in FLOPs . The number of channels ($N$) and kernel size ($k_1$) in the hidden layer are more critical in the trade-off between decoding FLOPs and R-D performance, and we leave a more detailed architecture search to future work.

\section{Related works}

\paragraph{Computationally efficient neural compression}

To reduce the high decoding complexity in neural compression, 
Johnston et al. \cite{johnston2019computationally} proposed to prune out filters in convolutional decoders with group-Lasso regularization \cite{gordon2018morphnet}.
Rippel and Bourdev \cite{rippel2017waveone} developed one of the earliest lossy neural image codecs with comparable running time to classical codecs, based on a multi-scale autoencoder architecture inspired by wavelet transforms such as JPEG 2000. 
Recent works propose computationally efficient neural compression architectures based on residual blocks \cite{cheng2020learned, he2022elic}, more lightweight entropy model \cite{he2021checkerboard, he2022elic}, network distillation \cite{wang2023EVC}, and accelerating learned entropy coding \cite{liu2021lossless}. We note there is also related effort on improving the compression performance of traditional codecs with learned components while maintaining high computational efficiency \cite{duong2023multi, isik2023sandwiched}.

\paragraph{Test-time / encoding optimization in compression}
The idea of improving compression performance with a powerful, content-adaptive encoding procedure is well-established in  data compression.
Indeed, vector quantization \cite{gersho2012vector} can be seen as implementing the most basic and general form of an optimization-based encoding procedure, and can be shown to be asymptotically optimal in rate-distortion performance \cite{cover1999elements}.
The encoders in commonly used traditional codecs such as H.264 \cite{sullivan2004h} and HEVC \cite{sze2014high} are also equipped with an exhaustive search procedure to select the optimal block partitioning and coding modes for each image frame. 
More recently, the idea of iterative and optimization-based encoding is becoming increasingly prominent in nonlinear transform coding \cite{campos2019content, yang2020improving, vanrozendaal2021overfitting}, as well as computer vision in the form of implicit neural representations \cite{park2019deepsdf, mildenhall2021nerf}. 
It is therefore interesting to see whether ideas from vector quantization and implicit neural representations may prove fruitful for further reducing the decoding complexity in NTC.

\paragraph{Manifold/metric learning}
A distantly related line of work is in metric learning with deep generative models, where the idea is to learn a latent representation of the data such that distance in the latent space preserves the similarity in the data space. 
Chen et al. \cite{chen2018metrics} proposes the use of the Riemannian distance metric induced by a decoding transform of a latent variable to measure similarity in the data space. 
Further, they proposed to learn flat manifolds with VAEs \cite{chen2020learning}, whose decoder essentially captures the geodesic distance between data points in terms of the Euclidean distance between their representations in the latent space. Their method is based on regularizing the Jacobian $\mathbf{J}$ of the decoder such that $\mathbf{J}^T \mathbf{J} \propto \mathbf{I}$, resulting in a length-preserving decoder and a latent space with low curvature, similar to what we observe with learned synthesis transforms in Section~\ref{sec:decoder-manifold-study}.

\section{Discussion}

In this work, we took a step towards closing the enormous gap between the decoding complexity of neural and traditional image compression methods. 
The main idea is to exploit the often asymmetrical computation budget of encoding and decoding: by pairing a lightweight decoder with 
a powerful encoder, we can obtain high R-D performance while enjoying low decoding complexity. We formalize this intuition theoretically, and show that the encoding procedure affects the R-D cost of lossy compression via an inference gap, and more powerful encoders improve R-D performance by reducing this gap.
In our implementation, we adopt shallow decoding transforms inspired by classical codecs such as JPEG and JPEG 2000, while employing more sophisticated encoding methods including iterative inference.
Empirically, we show that by pairing a powerful encoder with a shallow decoding transform, the resulting method achieves R-D performance competitive with BPG and the base Mean-Scale Hyperprior architecture \cite{minnen2018joint}, while reducing the complexity of the synthesis transform by over an order of magnitude. %
We suspect that the synthesis complexity can be further reduced by going beyond the transposed convolutions used in this work, e.g., via sub-pixel convolution \cite{shi2016real} or (transposed) depthwise convolution, as well as by exploiting the sparsity \cite{mallat2008wavelet} of the transform coefficients especially at low bit-rates. 

The success of nonlinear transform coding \cite{balle2021ntc} over traditional transform coding can be mostly attributed to (1) data-adaptive transforms and (2) expressive deep entropy models. %
We focused on improving the R-D-Compute efficiency of the synthesis transform, given that it accounts for the vast majority of decoding complexity in existing approaches, and left the hyperprior \cite{minnen2018joint} unchanged.
As a result, entropy decoding (via the hyper-synthesis transform) now takes up a majority (50 \% - 80\%) of the overall decoding computation in our method. Interestingly, we note that related work on flat manifolds found it necessary to use an expressive prior to learn a distance-preserving decoding transform \cite{chen2020learning}, and recent work in video compression \cite{mentzer2022vct} also features a simplified transform in the data space and a more expressive and computationally expensive entropy model.
Given recent advances in computationally efficient entropy models \cite{he2022elic, he2021checkerboard, liu2021lossless}, we are optimistic that the entropy decoder in our approach  can be significantly improved in rate-distortion-complexity, and leave this important direction to future work.

A limitation of our shallow synthesis is its worse performance on perceptual distortion compared to deeper architectures. Our study focused on the MSE distortion as in traditional transform coding; in this setting, it is known that an orthogonal linear transform gives optimal R-D performance for Gaussian-distributed data \cite{gersho2012vector}. However, the distribution of natural images is far from Gaussian, and compression methods are increasingly evaluated on perceptual metrics such as MS-SSIM \cite{wang2003msssim} --- both factors motivating the use of nonlinear transforms. 
We believe insights from signal processing and deep generative modeling may inspire more efficient nonlinear transforms with high perceptual quality, or an efficient pipeline based on a cheap MSE-optimized reconstruction followed by generative artifact removal/denoising for good perceptual quality.

\section*{Acknowledgements}
Yibo Yang acknowledges support from the HPI Research Center in Machine Learning and Data Science at UC Irvine. Stephan Mandt acknowledges support by the National Science Foundation (NSF) under an NSF CAREER Award IIS-2047418 and IIS-2007719. Stephan Mandt thanks Qualcomm for unrestricted research gifts. 
We thank David Minnen for valuable feedback and suggestions.

{\small
\bibliographystyle{ieee_fullname}
\bibliography{egbib}
}

\newpage
\onecolumn

\section{Appendix to ``Computationally-Efficient Neural Image Compression with Shallow Decoders''}\blfootnote{Our code can be found at \url{https://github.com/mandt-lab/shallow-ntc}.}

\subsection{More details on the two-layer synthesis with residual connection}\label{app:2layer-res-arch-details}

\begin{figure}[ht]
    \centering
    \includegraphics[width=0.4\textwidth]{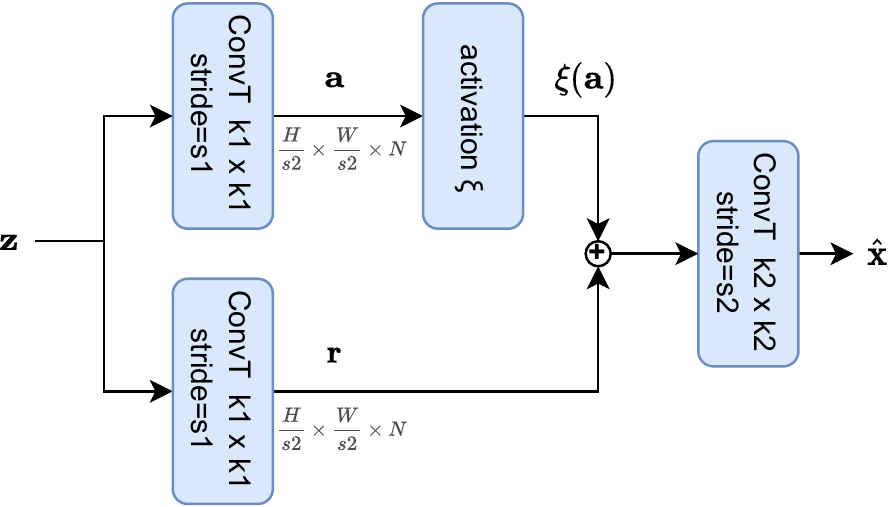}
    \caption{Diagram of the proposed two-layer synthesis transform.}
    \label{fig:2layer_arch}
\end{figure}

Fig.~\ref{fig:2layer_arch} illustrates the proposed two-layer architecture with a residual connection, where $\a = \texttt{conv\_1}(\z)$ denotes the output of the first transposed conv layer. 
We implement the residual connection (the lower computation path in the figure) with another transposed convolution layer $\texttt{conv\_res}$, using the same configuration (stride, kernel size, etc.) as $\texttt{conv\_1}$. 
In our main experiments we use $k_1=13, s_1=8, k_2=5, s_2=2$ and $N=12$.

The residual connection $\br$ is inspired by its success in recent NTC architectures \cite{cheng2020learned, he2022elic}, and can also be interpreted as a data-dependent and spatially-varying bias term that modulates the nonlinear activation $\xi(\a)$. At lower bit-rates, we found employing the residual connection to be more parameter- and compute-efficient than a simple composition of two transposed layers without the residual connection.

In further experiments, we re-trained models with ``mixed quantization'' \cite{minnen2020channel} instead of additive uniform noise as in the main paper, and found that with comparable decoding complexity, a simple two-layer architecture with $N=24$ hidden channels (and no residual connection) in fact slightly outperforms the one with residual connection and $N=12$, while keeping all other hyperparameters the same.

\subsection{Theoretical Result} \label{app:rd-cost-decomposition}
In the following, we derive the decomposition of the R-D cost of neural lossy compression. 
To lighten notation, we use non-bold letters ($x, z$ instead of $\x, \z$), and adopt the general setting where the latent space $\setZ$ is a Polish space (which includes, among many other examples, the  Euclidean space $\mathbb{R}^{|\setZ|}$ commonly used for continuous latent variables, or the set of lattice points $\mathbb{R}^{|\setZ|}$ in nonlinear transform coding), and $P_Z$ is a prior probability measure.
We present results in terms of measures for generality, but for readers unfamiliar with measure theory it is harmless to focus on the common case where
$P_Z$ admits a density $p(z)$ (denoted $p(\z)$ in the main paper), s.t., $P_Z(dz) = p(z) dz$; in the discrete case, $p(z)$ is a PMF, and the integral (w.r.t. the counting measure on $\setZ$) reduces to a sum. Similarly, $Q_{Z|X}$ is family of probability measures, such that for each value of $x$ it defines a conditional distribution $Q_{Z|X=x}$, which may admit a density $q(z|x)$. %

A (learned) lossy compression codec consists of a prior distribution $P_Z$, a stochastic encoding transform $Q_{Z|X}$, and a deterministic decoding transform $g: \setZ \to \sethX$.
Suppose relative entropy coding \cite{cuff2008, theis2021algorithms, flamich2020cwq} operates with minimal overhead, i.e.,  a sample from $Q_{Z|X=x}$ can be transmitted under $P_Z$ with a bit-rate close to $KL(Q_{Z|X=x} \| P_Z)$ (which may require us to perform block coding), then 
given a data realization $x$, the rate-distortion compression cost is, on average,
\begin{align}
\mathcal{L}(Q_{Z|X}, g, P_Z, x) := \lambda \E_{z \sim Q_{Z|X=x}} [\rho(x, g(z))] + KL(Q_{Z|X=x} \| P_Z), \label{eq:per-instance-nelbo}
\end{align} 
where $\rho: \mathcal{X} \times \sethX \to [0, \infty)$ is the distortion function, and $\lambda \geq 0$ is a fixed  hyperparameter trading off between rate and distortion, and both $\rho$ and $\lambda$  are specified by the lossy compression problem in advance.

By Lemma 8.5 of \cite{gray2011entropy}, this compression cost admits a similar decomposition to the negative ELBO,

\begin{align}
    \mathcal{L}(Q_{Z|X}, g, P_Z, x) 
   =   - \log \Gamma_{g, P_Z} (x) + KL(Q_{Z|X=x} \| P_{Z|X=x}), \label{eq:gibbs}
\end{align}
where $P_{Z|X}$ denotes the Markov kernel (transitional distribution) defined by
\begin{align}
    P_{Z|X=x}(dz) := \frac{e^{-\lambda \rho(x, g(z))} P_Z(dz)}{\Gamma_{g, P_Z}(x) },
\end{align}
and the normalizing constant is
\begin{align}
    \Gamma_{g, P_Z}(x) := \int_{\setZ} e^{-\lambda \rho(x, g(z))} P_Z(dz).
\end{align}
As in variational Bayesian inference, the normalizing constant has the interpretation of a marginal log-likelihood specified by the prior $P_Z$ and model $g$. 
We note that the definition of $P_{Z|X}$ depends on $g$ and $P_Z$, but leave this out to lighten notation. 
Eq.~\ref{eq:gibbs} together with the non-negativity of KL divergence imply that $P_{Z|X}$ is the optimal channel (``inference distribution'') for reverse channel coding, under which the compression cost equals:
\begin{align}
    \min_{Q_{Z|X=x}} \lambda \E_{z \sim Q_{Z|X=x}} [\rho(x, g(z))] + KL(Q_{Z|X=x} \| P_Z) = -\log \Gamma_{g, P_Z}(x) \label{eq:opt-rec-channel}
\end{align}

Let $P_X$ be the data distribution. Taking the expected value of Eq.~\ref{eq:per-instance-nelbo} w.r.t. $x \sim P_X$ gives the population-level compression cost, which can then be rewritten as follows
\begin{align}
\mathcal{L}(Q_{Z|X}, g, P_Z) &:= 
\lambda \E_{P_X Q_{Z|X}} [\rho(X, g(Z))] + \E_{x \sim P_X}[KL(Q_{Z|X=x} \| P_Z)] \\
&= \E_{x\sim P_X}[- \log \Gamma_{g, P_Z} (x)] + \E_{x\sim P_X}[KL(Q_{Z|X=x} \| P_{Z|X=x})] \\
&= \inf_{P'_Z, \omega}\E_{x\sim P_X}[- \log \Gamma_{\omega, P'_Z} (x)] + \left(\E_{x\sim P_X}[- \log \Gamma_{g, P_Z} (x)] -\inf_{P'_Z, \omega}\E_{x\sim P_X}[- \log \Gamma_{\omega, P'_Z} (x)] \right) \\ &+ \E_{x\sim P_X}[KL(Q_{Z|X=x} \| P_{Z|X=x})] \\
&= \inf_{\omega \in \mathcal{G}} F_{\omega}(\lambda) + \underbrace{\left(\E_{x\sim P_X}[- \log \Gamma_{g, P_Z} (x)] - \inf_{\omega \in \mathcal{G}}  F_{\omega}(\lambda) \right)}_{\text{modeling gap}} + \underbrace{\E_{x\sim P_X}[KL(Q_{Z|X=x} \| P_{Z|X=x})]}_{\text{inference gap}}, 
\end{align}
where for any choice of decoder function $\omega \in \mathcal{G}$, we define
\begin{align}
    F_{\omega}(\lambda):= \inf_{Q_{Z|X}} I(X; Z) + \lambda \E[\rho(X, \omega(Z))]
\end{align}
as the $R$-axis intercept of the line tangent to the $\omega$-dependent rate-distortion function \cite{yang2022towards} \footnote{The $g$-dependent distortion function \cite{yang2022towards} is defined as $R_g(D):= \inf_{Q_{Z|X}: \E[\rho(X, g(Z))] \leq D} I(X; Z)$.} with slope $-\lambda$. The last equation follows from Eq.~\ref{eq:opt-rec-channel} and the following calculation
\begin{align}
    & \inf_{P'_Z, \omega} \E_{x\sim P_X}[- \log \Gamma_{\omega, P'_Z} (x)] \\
    &= \inf_{P'_Z, \omega} \inf_{Q_{Z|X}} \lambda \E_{P_X Q_{Z|X}} [\rho(X, \omega(Z))] + \E_{x \sim P_X}[KL(Q_{Z|X=x} \| P'_Z)]  \\
     &= \inf_{\omega} \inf_{Q_{Z|X}} \inf_{P'_Z} \lambda \E_{P_X Q_{Z|X}} [\rho(X, \omega(Z))] + \E_{x \sim P_X}[KL(Q_{Z|X=x} \| P'_Z)]  \\
    &= \inf_{\omega} \inf_{Q_{Z|X}} \lambda \E_{P_X Q_{Z|X}} [\rho(X, \omega(Z))] + \inf_{P'_Z} 
 \E_{x \sim P_X}[KL(Q_{Z|X=x} \| P'_Z)]  \\
     &= \inf_{\omega} \inf_{Q_{Z|X}} \lambda \E_{P_X Q_{Z|X}} [\rho(X, \omega(Z))] +  I(P_X Q_{Z|X}) \\
     &= \inf_{\omega} F_\omega(\lambda)
\end{align}

To summarize, we have broken down the R-D cost for a data source $P_X$ into three terms,
\begin{align}
\mathcal{L}(Q_{Z|X}, \omega, P_Z) =\inf_{\omega \in \mathcal{G}} F_\omega(\lambda) + \underbrace{\left(\E_{x\sim P_X}[- \log \Gamma_{g, P_Z} (x)] - \inf_{\omega \in \mathcal{G}}  F_\omega(\lambda) \right)}_{\text{modeling gap}} + \underbrace{\E_{x\sim P_X}[KL(Q_{Z|X=x} \| P_{Z|X=x})]}_{\text{inference gap}}.
\end{align}
\begin{itemize}
    \item The first term (which we dentoed by the shorthand ``$\mathcal{F}(\mathcal{G})$'' in the main text) represents the fundamentally irreducible cost of compression determined by the source $P_X$ and the family $\mathcal{G}$ of decoding transforms used. This is the information-theoretically optimal cost of compression within the transform family $\mathcal{G}$.
    If we let $F(\lambda):= \inf_{Q_{\hat X|X}} I(X; \hat X) + \lambda \E[\rho(X, \hat X)]$ be the optimal Lagrangian associated with the R-D function of $P_X$, then it can be shown that  \cite{yang2022towards} $F(\lambda) \leq F_g(\lambda)$. When the latent space $\mathcal{Z}$ and the transform family $\mathcal{G}$ are sufficiently large, it holds that $F(\lambda) = \inf_g F_g(\lambda)$, i.e., the first term of the R-D cost is determined solely by the rate-distortion function of the source distribution $P_X$ (and distortion function $\rho$), which is the lossy analogue of the Shannon entropy.
    \item The second term represents the excess cost of doing compression with a \emph{particular} transform $g$ and prior $P_Z$  compared to the \emph{best possible} transform and prior, while always operating with the optimal channel (Eq.~\ref{eq:gibbs}) in each case. Note that this term only depends on the modeling choices of $g$ and $P_Z$, and does not depend on the encoding/inference distribution $Q_{Z|X}$; therefore we call it the \emph{modeling gap}. It is due largely to imperfect model training/optimization, and/or a mismatch between the training data and the target data $P_X$ (which may not be the same).
    \item The third term represents the overhead of compression caused by a (potentially) sub-optimal encoding/inference distribution $Q_{Z|X}$, given a particular model $g$ and $P_Z$. This overhead can be eliminated by using the optimal channel $P_{Z|X}$ given in Eq.~\ref{eq:gibbs} (which depends on $g$ and $P_Z$). Therefore we call this term the \emph{inference gap}.
\end{itemize}

\paragraph{Remarks} The decomposition of the lossy compression cost has a natural parallel to that of lossless compression under a latent variable model. 

Consider a latent variable model $(p_\theta(z), p_\theta(x|z))$ with parameter vector $\theta$, which defines a model of the marginal data density $p_\theta(x):= \int_\mathcal{Z}  p_\theta(x|z) p_\theta(z) dz$.
The cost of lossless compression under ideal bits-back coding \cite{frey1998bayesian, townsend2019practical} is equal to the negative ELBO, and admits a similar decomposition \cite{zhang2022generalization}:
\begin{align}
    & \E_{x \sim P_X}[ \E_{z \sim q(z|x)}[-\log p_\theta(x|z)] + KL(q(z|x)\|p_\theta(z)) ]\\
    = & \E_{x\sim P_X}[-\log p_\theta(x)] + \E_{x\sim P_X}[KL(q(z|x)\|p_\theta(z|x))] \\
= & \underbrace{H[P_X]}_{\text{data entropy}} + \underbrace{ KL(P_X \|p_\theta(x))}_{\text{modeling gap}} + \underbrace{\E_{x\sim P_X}[KL(q(z|x)\|p_\theta(z|x))]}_{\text{inference gap}}
\end{align}

Again, we have decomposed the compression cost into a first term that represents the intrinsic compressibility of the data, a second term that depends entirely on the choice of the model, and a third overhead term from using a sub-optimal inference distribution $q(z|x)$ and which can be eliminated using the optimal inference distribution $p_\theta(z|x) \propto p_\theta(x|z) p_\theta(z)$ (the Bayesian posterior) given each choice of our model.

\newpage
\subsection{Implementation and reproducibility details}
Our models are implemented in tensorflow using the \texttt{tensorflow-compression}\footnote{\url{https://github.com/tensorflow/compression}} library.
We implemented the Mean-Scale Hyperprior model based on the open source code \footnote{\url{https://github.com/tensorflow/compression/blob/master/models/bmshj2018.py}} and architecture details from Johnston et al. \cite{johnston2019computationally}. We borrowed the ELIC \cite{he2022elic} transforms from the VCT repo \footnote{\url{https://github.com/google-research/google-research/blob/master/vct/src/elic.py}}.

Our experiments were run on Titan RTX GPUs. All the models were trained with the Adam optimizer following standard procedure (e.g., in \cite{minnen2018joint}) for a maximum of 2 million steps. We use an initial learning rate of $1e-4$, then decay it to $1e-5$ towards the end of training. For each model architecture, we trained separate models for $\lambda \in \{0.00125, 0.0025, 0.005, 0.01, 0.02, 0.04, 0.08\}$. 
For SGA \cite{yang2020improving}, we use similar hyperparameters as in \cite{yang2020improving}, using Adam optimizer, learning rate $5e-3$, and a temperature schedule of $\tau(t) = 0.5 \exp\{ - 0.0005 (\max(0, t - 200)) \} $ for 3000 gradient steps.

\subsection{Additional Results} \label{app:additional_results}

\subsubsection{Results on the reconstruction manifold.}\label{app:latent_traversal}

We train three popular NTC architectures \cite{balle2018hyper, minnen2018joint, minnen2020channel} for MSE distortion with $\lambda=0.08$ and observe similar results when traversing the manifold of reconstructed images. We use random pairs of image crops from COCO \cite{lin2014microsoftcoco} to define the start and end points of latent traversal (see Sec.~\ref{sec:decoder-manifold-study}); we use the same random seed across the three different architectures. All the images are scaled to $[-0.5, 0.5]$.

\begin{figure}[h]
     \centering
     \begin{subfigure}[b]{0.3\textwidth}
         \centering
         \includegraphics[width=\textwidth]{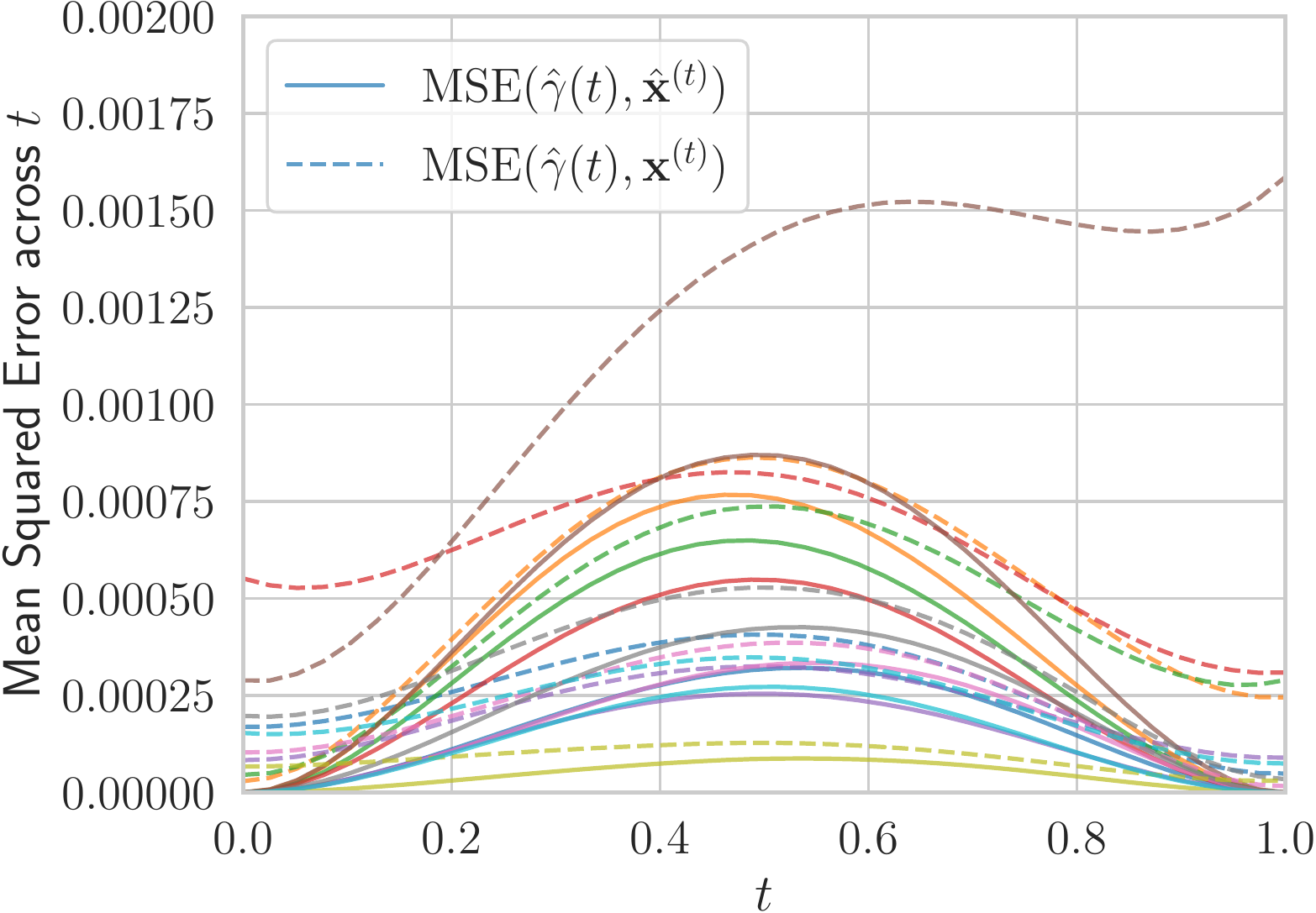}
         \caption{Factorized Prior \cite{balle2018hyper}.}
     \end{subfigure}
     \hfill
     \begin{subfigure}[b]{0.3\textwidth}
         \centering
         \includegraphics[width=\textwidth]{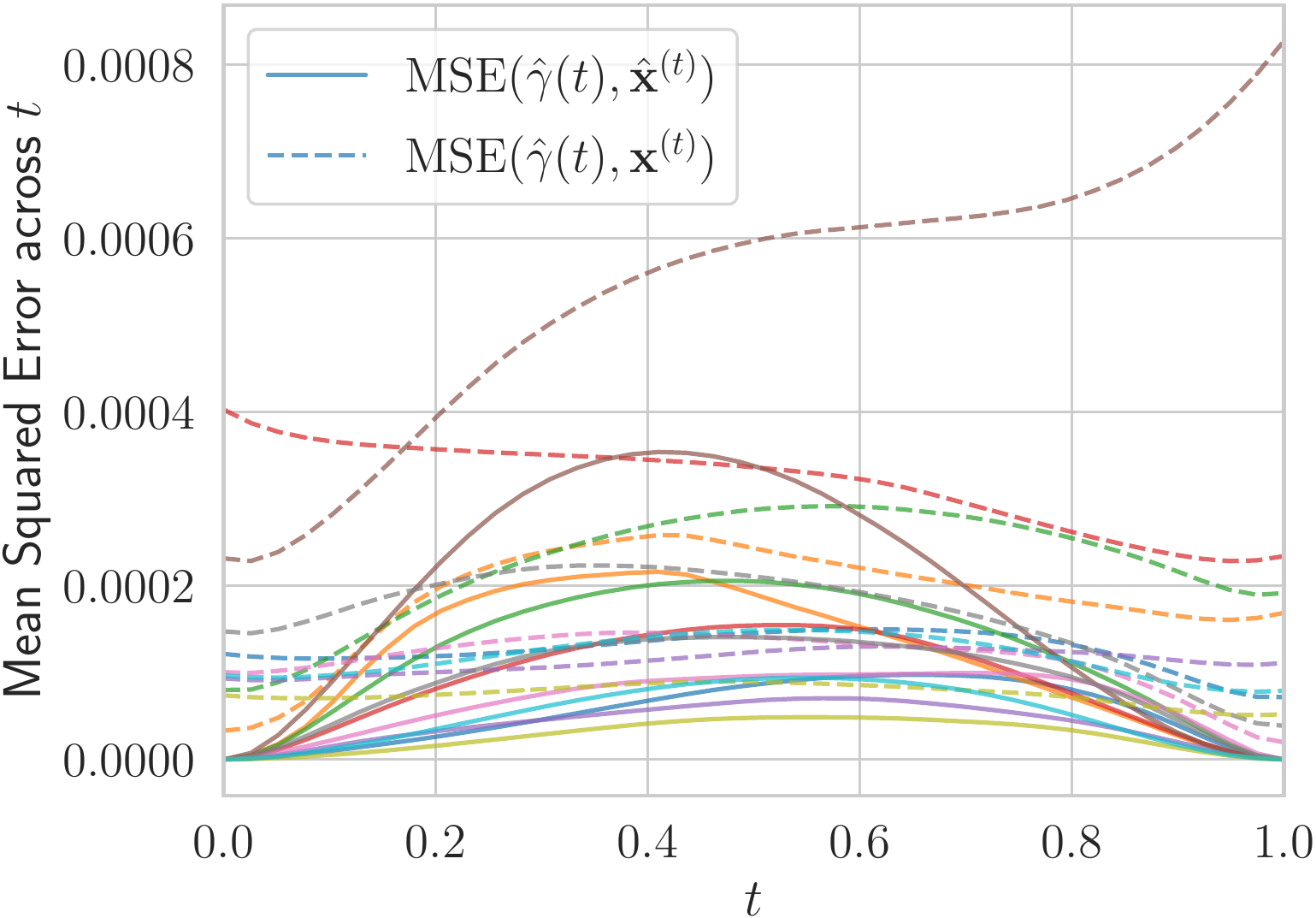}
         \caption{Mean-Scale Hyperprior \cite{minnen2018joint}.}         %
     \end{subfigure}
     \hfill
     \begin{subfigure}[b]{0.3\textwidth}
         \centering
         \includegraphics[width=\textwidth]{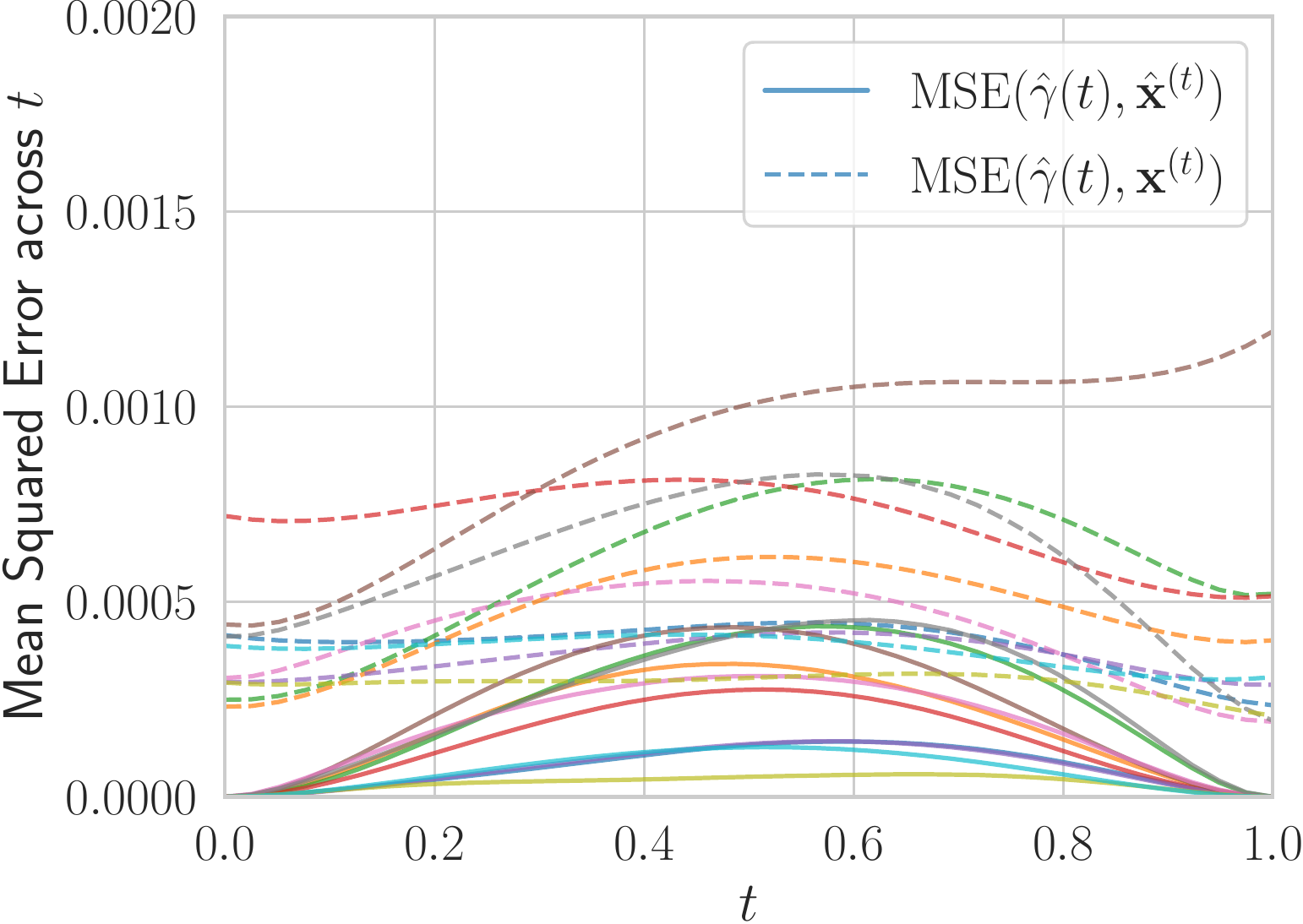}
         \caption{Channel-wise Autoregressive \cite{minnen2020channel}. }
     \end{subfigure}
        \caption{The distance from the trajectory of decoded curve $\hat\gamma(t)$ to the straight path between end-point reconstructions $\hatx^{(t)}:=(1-t) \hatx^{(0)} + t \hatx^{(1)}$, and to the straight path between ground truth images $\hatx^{(t)} := (1-t) \x^{(0)} + t \x^{(1)}.$ }
        \label{fig:more-manifold-mses}
\end{figure}

\paragraph{MSEs between trajectories}
In Fig.~\ref{fig:more-manifold-mses} shows the distance between the resulting trajectory of decoded curve $\hat\gamma(t)$ and two kinds of straight paths, interpolating between reconstructions $\hatx^{(t)}:=(1-t) \hatx^{(0)} + t \hatx^{(1)}$ or ground truth images $\hatx^{(t)} := (1-t) \x^{(0)} + t \x^{(1)}.$ See detailed discussions in Sec.~\ref{sec:decoder-manifold-study}.

\paragraph{Quantifying the curvature of decoded curves of reconstructions.}
Additionally, we quantify how much the curves of reconstructed images deviate from straight paths by computing the curve lengths. 
Recall given two tensors of latent coefficients $(\z^{(0)},  \z^{(1)})$ (obtained by passing two images $(\x^{(0)}, \x^{(1)})$ through the analysis transform), we let $\gamma: [0, 1] \to \setZ$ be the straight line in the latent space defined by their convex combination, i.e., $\gamma(t) := (1-t) \z^{(0)} + t \z^{(1)}$. The curve of reconstructions is then defined by $\hat \gamma(t) := g(\gamma(t))$, with end-points $\hatx^{(0)} := g(\z^{(0)})$ and $\hatx^{(1)} := g(\z^{(1)})$.

Following Chen et al. \cite{chen2018metrics}, the curve length of $\hat \gamma$ is given by

\begin{align}
    L(\hat \gamma) &:= \int_0^1 \| \frac{\partial g(\gamma(t))} {\partial t} \| \text{d} t = \int_0^1 \| \frac{\partial g(\gamma(t))} {\partial \gamma(t)} \frac{\partial \gamma(t)} {\partial t} \| \text{d} t = \int_0^1 \| \mathbf{J}_t \mathbf{v} \| \text{d} t  = \int_0^1 \sqrt{ \mathbf{v}^T {\mathbf{J}_t}^T \mathbf{J}_t \mathbf{v} } ~ \text{d} t \label{eq:curve-length}
\end{align}
where $\mathbf{J}_t \in \mathbb{R}^{|\setX| \times |\setZ|}$ is the Jacobian of the synthesis transform evaluated at $\z^{(t)} = \gamma(t)$, and $\mathbf{v} = \frac{\partial \gamma(t)} {\partial t} = \z^{(1)} - \z^{(0)}$ is the (constant) curve velocity. As in \cite{chen2018metrics}, we compute this integral approximately with a Riemann sum.

The shortest path (in Euclidean geometry) between the two end-points of $\hat \gamma$ is given simply by the linear interpolation $(1-t) \hatx^{(0)} + t \hatx^{(1)} $, with a distance of $\|\hatx^{(1)} - \hatx^{(0)}\|$.
Therefore, we define the curve-to-shortest-path length ratio 
\begin{align}
    \eta := \frac{L(\hat \gamma)} {\|\hatx^{(1)} - \hatx^{(0)}\|}
\end{align}
as a measure of how much the curve $\hat \gamma$ deviates from a straight path, with $\eta = 1$ indicating a completely straight line.

\begin{figure}[t]
     \centering
     \begin{subfigure}[b]{0.3\textwidth}
         \centering
         \includegraphics[width=\textwidth]{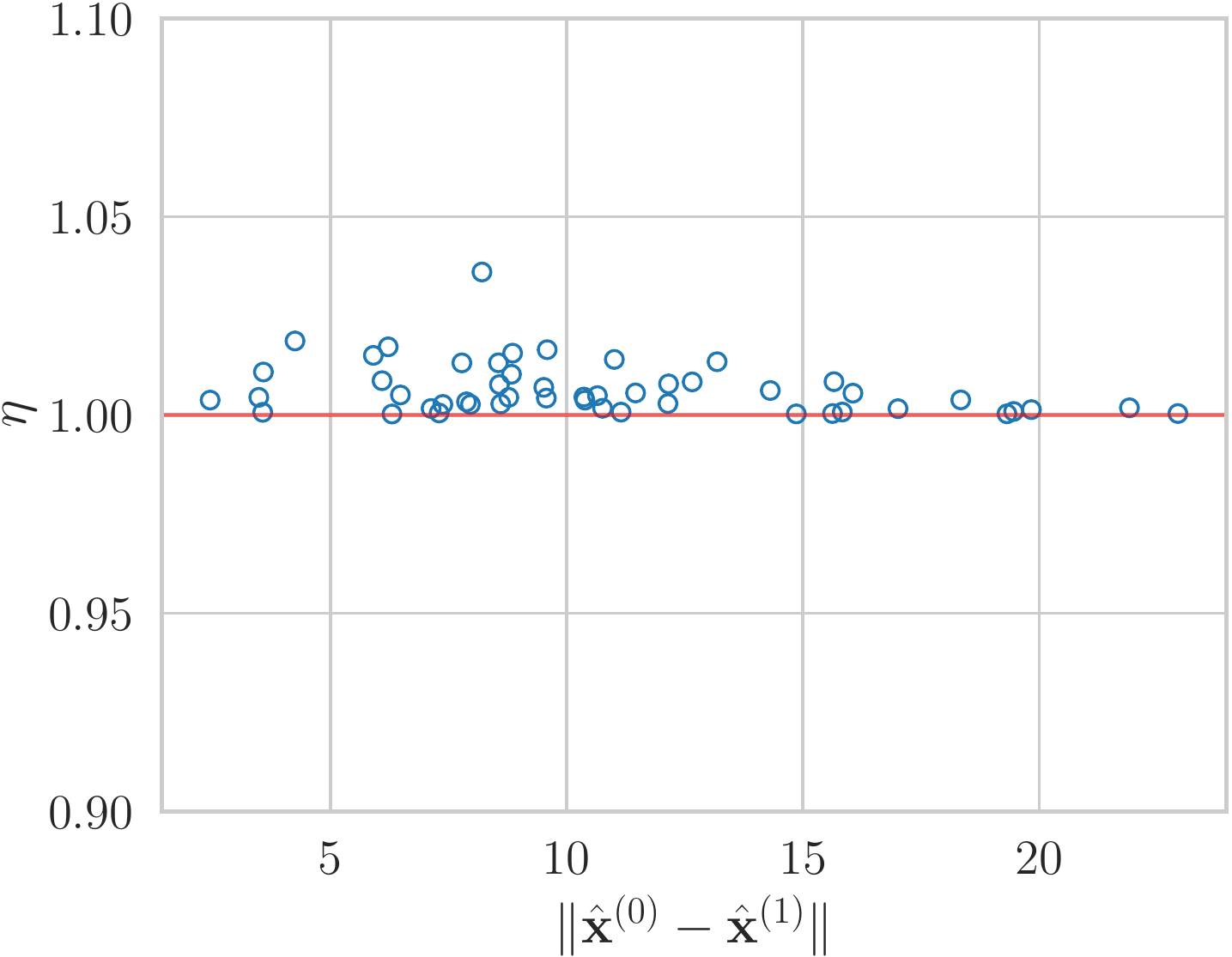}
         \caption{Factorized Prior \cite{balle2018hyper}.}
     \end{subfigure}
     \hfill
     \begin{subfigure}[b]{0.3\textwidth}
         \centering
         \includegraphics[width=\textwidth]{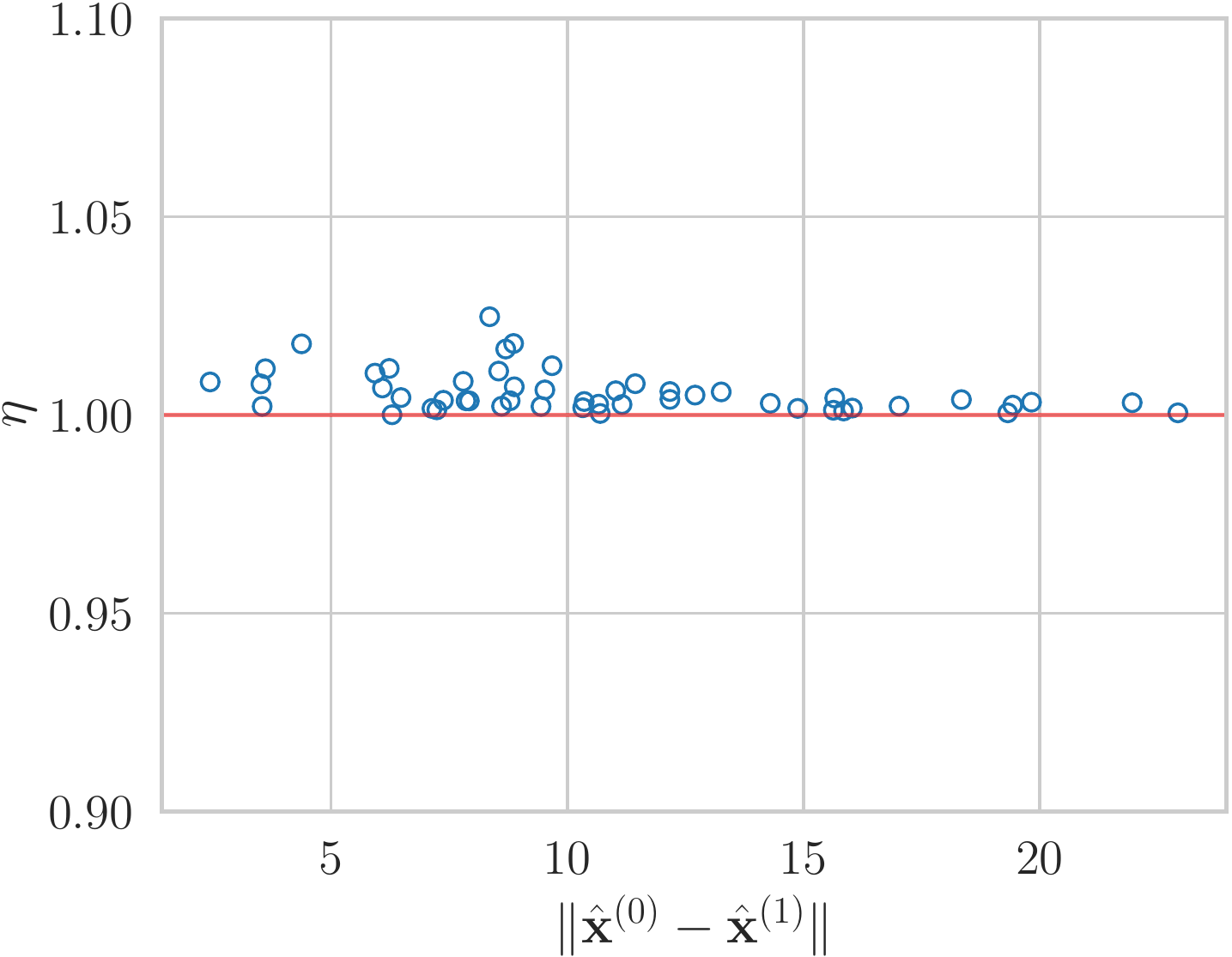}
         \caption{Mean-Scale Hyperprior \cite{minnen2018joint}.}         %
     \end{subfigure}
     \hfill
     \begin{subfigure}[b]{0.3\textwidth}
         \centering
         \includegraphics[width=\textwidth]{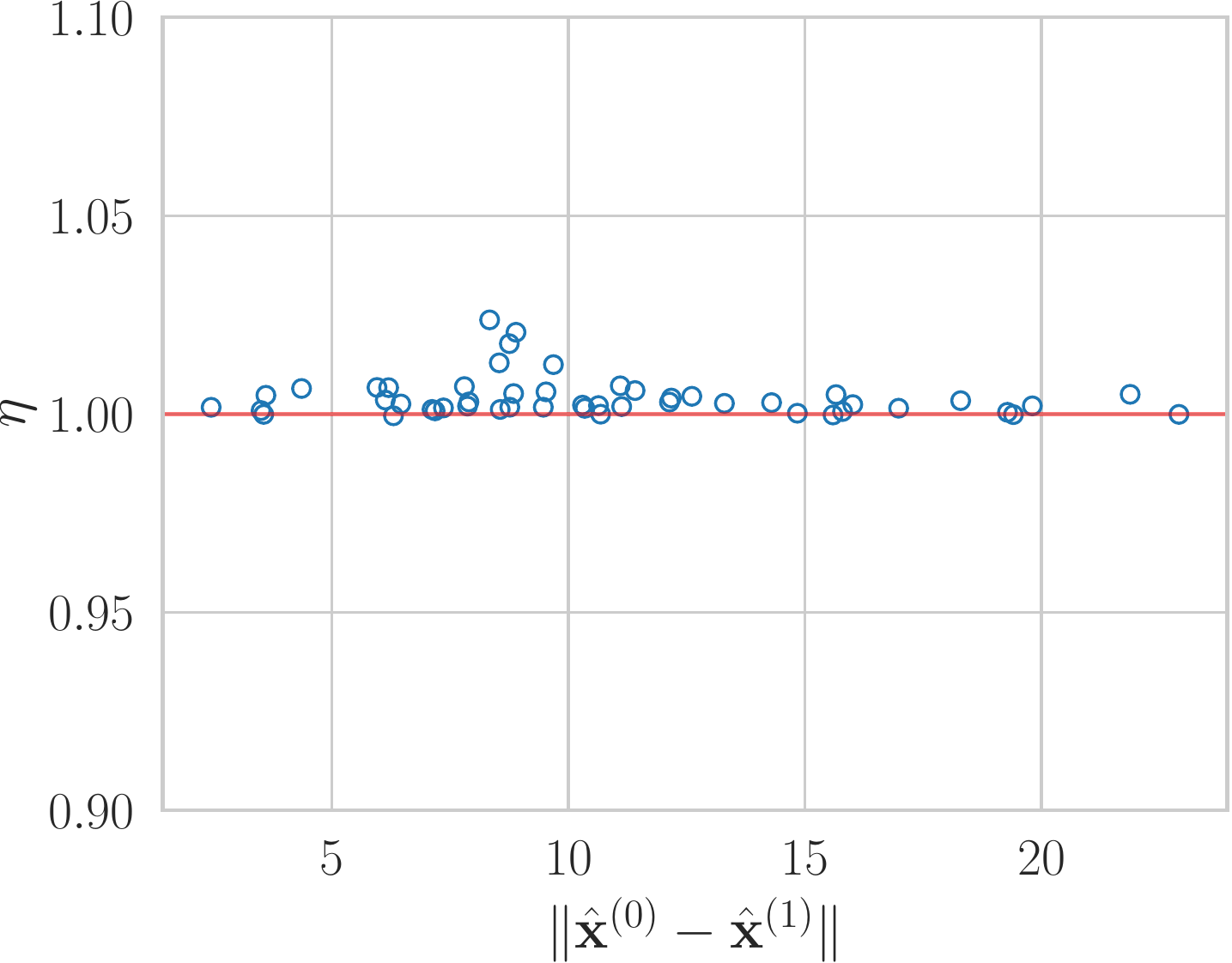}
         \caption{Channel-wise Autoregressive \cite{minnen2020channel}. }
     \end{subfigure}
        \caption{The curve-length ratio $\eta$ v.s. the straight-path-length for randomly chosen image pairs, in three different nonlinear transform coding architectures \cite{balle2018hyper, minnen2018joint, minnen2020channel}. In all cases, the curve lengths are close to the lengths of straight paths.}
        \label{fig:eta-scatter}
\end{figure}

We compute the curve-to-shortest-path-length ratio $\eta$ on 50 randomly chosen image pairs in three NTC architectures \cite{balle2018hyper, minnen2018joint, minnen2020channel}. 
We use random $16 \times 16$ image crops (the results are similar for larger images) from COCO \cite{lin2014microsoftcoco}. We compute the curve length integral in Eq.~\ref{eq:curve-length} via a Riemann sum, $\frac{1}{T} \sum_{t_i} \| \mathbf{J}_{t_i} (\z^{(1)} - \z^{(0)})  \|  $, with $T=100$.
Fig.~\ref{fig:eta-scatter} plots the resulting $\eta$ values against the straight-path lengths. In all cases, the curve lengths are close to the straight-path lengths ($\eta$ concentrated near 1), and this property appears to hold globally across randomly chosen image pairs.

\subsubsection{Visualizing the filters of synthesis transforms}
\begin{figure*}[h!]
    \centering
    \begin{subfigure}[t]{1\textwidth}
        \centering
        \includegraphics[width=1\linewidth]{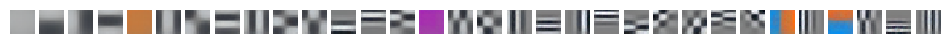}
        \caption{Top 20 filters of the Mean-Scale Hyperprior model, $\lambda=0.00125$, giving bpp=0.10, psnr=27.3 on Kodak.}
    \end{subfigure} \\
    \begin{subfigure}[t]{1\textwidth}
        \centering
        \includegraphics[width=1\linewidth]{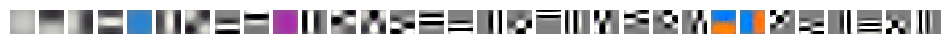}
        \caption{Top 20 filters of the two-layer synthesis model, $\lambda=0.00125$, giving bpp=0.12, psnr=27.3 on Kodak.}
    \end{subfigure} \\
    \begin{subfigure}[t]{1\textwidth}
    \centering
    \includegraphics[width=1\linewidth]{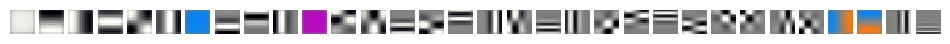}
    \caption{Top 20 filters of the JPEG-like synthesis model, $\lambda=0.00125$, giving bpp=0.12, psnr=27.0 on Kodak.}
    \end{subfigure} \\
    \begin{subfigure}[t]{1\textwidth}
        \centering
        \includegraphics[width=1\linewidth]{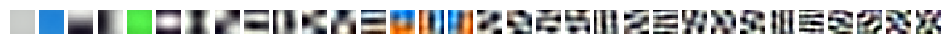}
        \caption{Principal components with the 20 largest eigenvalues, extracted from random 16x16 natural image patches.}
    \end{subfigure} 
    \caption{Visualization of learned filters in various neural compression methods with varying  synthesis transform complexity. The ``filters'' obtained from PCA are included for reference and show certain qualitative similarities. The filters are ordered by decreasing average bit-rate (a,b,c) or eigenvalue (d). See text description for more details.}\label{fig:filter-vis}
\end{figure*}
In Figure~\ref{fig:filter-vis}, we visualize the learned filters of the base Mean-Scale Hyperprior architecture (panel a), and the proposed architectures with two-layer synthesis (panel b) and JPEG-like synthesis (panel c) (both paired with ELIC's analysis transform). 
We also visualize the top 20 PCA components learned on 10000 random $16 \times 16$ color image patches from COCO (panel d).

For the neural compression methods, we visualize the synthesis filters corresponding to the 20 latent channels with the highest bit-rates on average (as determined on a small batch of validation images); following \cite{duan2022opening}, we produce the visualization for channel $i$ as follows: let $\mathbf{e}$ be a `basis' tensor of shape $[1, 1, C]$ (unit width/height) consisting of zeros except the $i$th channel, which equals 1; let $\mathbf{0}$ be a tensor of zeros with the same shape as $\mathbf{e}$; then the impulse response associated with channel $i$ is computed as $g(\delta \mathbf{e}) - g(\mathbf{0})$, where $\delta$ is a scaling factor which affects the color intensity when visualized. This results in a $16 \times 16$ colored image patch. We manually set a different $\delta$ for each architecture to result in a roughly comparable range of displayed colors, with $\delta \in [8, 20]$. We also apply a scaling factor when visualizing the principal components from PCA.

\subsubsection{Ablation results.} \label{app:more-ablations}

\begin{figure}[t]
     \centering
     \begin{subfigure}[b]{0.45\textwidth}
         \centering
         \includegraphics[width=\textwidth]{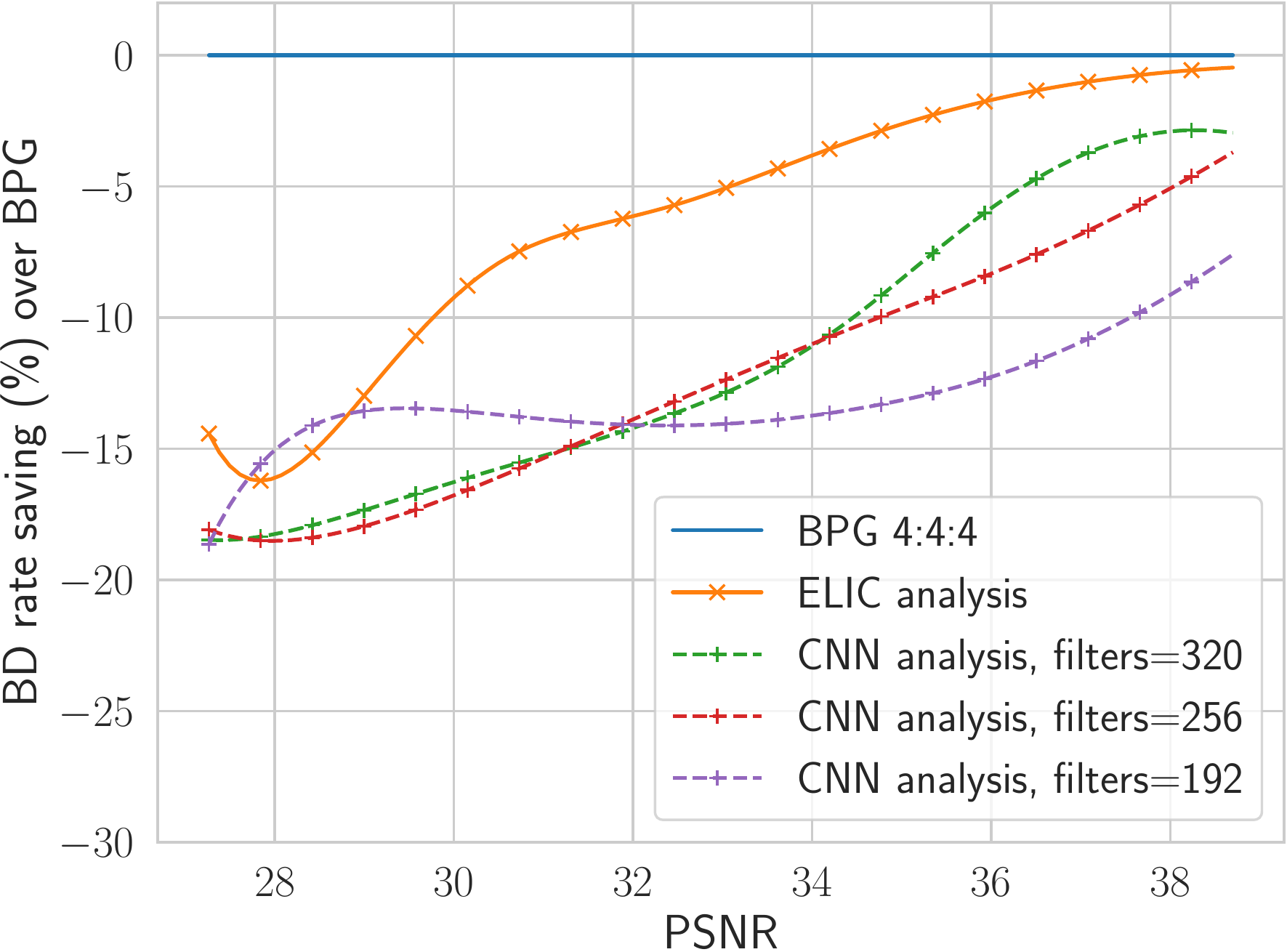}
         \caption{BD-rate savings over BPG on Kodak, for various choices of analysis transforms.}
     \end{subfigure}
     \hfill
     \begin{subfigure}[b]{0.45\textwidth}
         \centering
         \includegraphics[width=\textwidth]{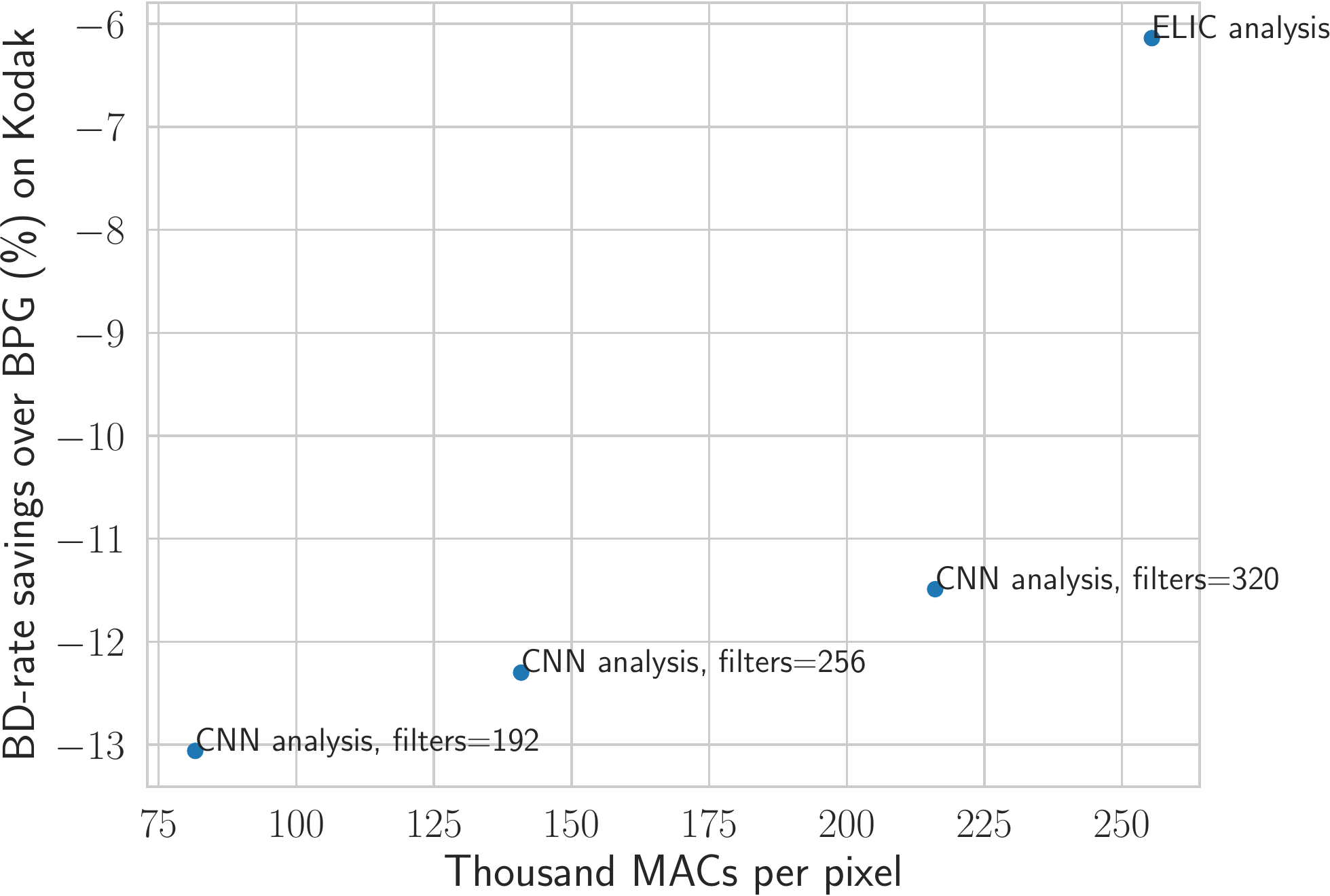}
         \caption{Aggregate BD-rate savings on Kodak, v.s. analysis transform complexity, measured in KMACs per pixel. }
     \end{subfigure}
        \caption{Ablation results on the choice of analysis transform in the proposed two-layer synthesis architecture. Using the CNN analysis from Mean-Scale Hyperprior \cite{minnen2018joint} gave relatively worse R-D performance than the analysis transform from ELIC \cite{he2022elic}. }
        \label{fig:comp-ana-rdc}
\end{figure}

\paragraph{The analysis transform.}
Here, we examine how different choices of the analysis transform affect the performance of our method based on the two-layer synthesis transform and ELIC's analysis transform. 

We adopt the simpler CNN analysis transform from Mean-Scale Hyperprior \cite{minnen2018joint}, which consists of 4 layers of convolutions with $F=192$ filters each, except for the last layer which outputs $C=320$ channels for the latent tensor. 
Fig.~\ref{fig:comp-ana-rdc} shows the resulting performance with varying $F$, in both BD-rate savings as well as computational complexity. We see that the CNN analysis gave worse performance than ELIC analysis, and the gap can be closed to some extent by increasing $F$, but with diminishing returns and increasingly high encoding complexity.

\begin{figure}[t]
    \centering
    \includegraphics[scale=0.5]{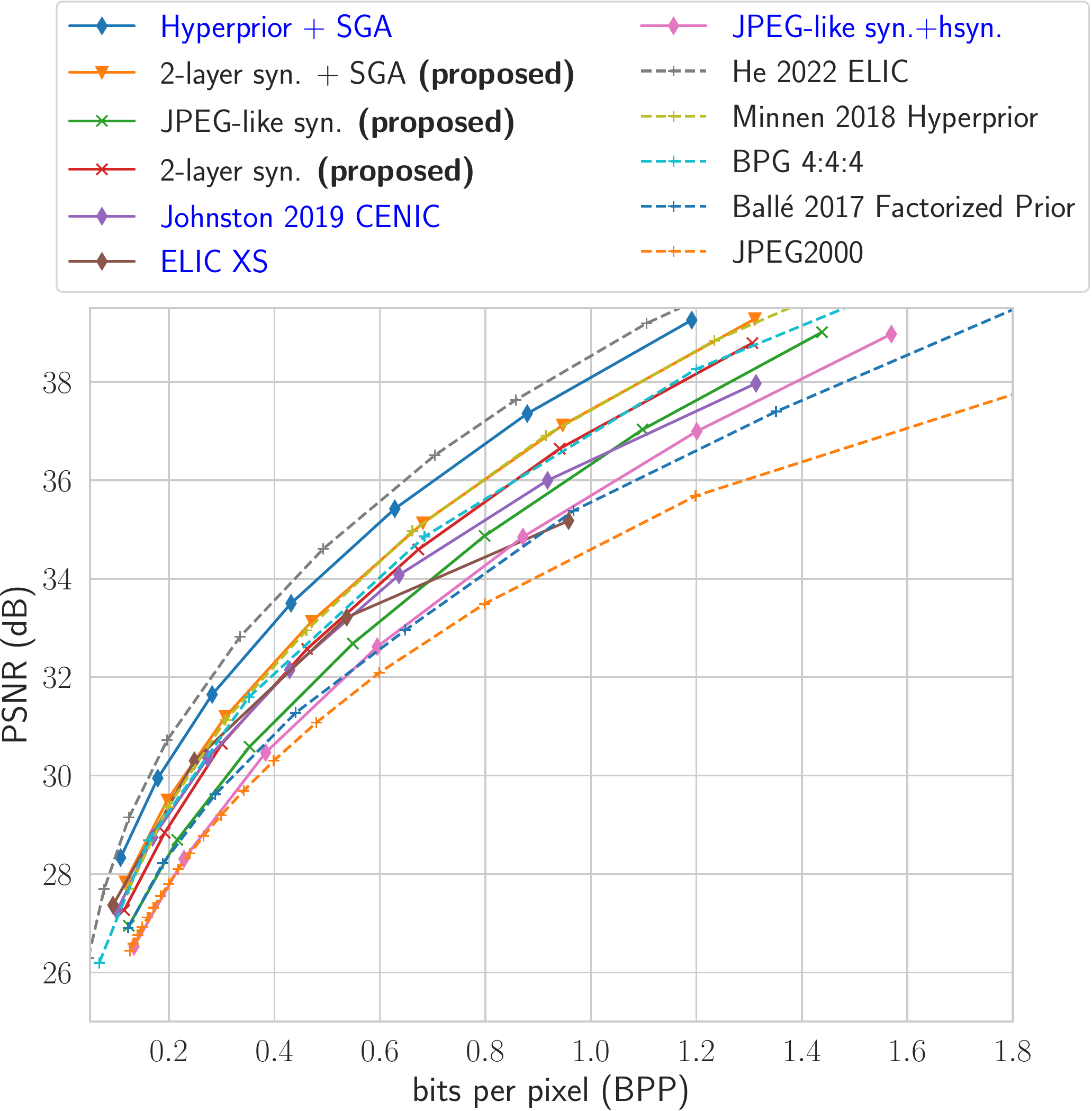}
    \caption{Miscellaneous additional results, with blue legend labels to be distinguished from results in the main paper. We (1) apply SGA also to the Hyperprior baseline; (2) simply scale down existing NTC architectures to roughly match the FLOPs of our two-layer synthesis; (3) explore a JPEG-like hyper synthesis transform to improve its computational efficiency. }
    \label{fig:additional-res}
\end{figure}

\paragraph{Additional investigations.}
We present results giving additional insight into our method and how it compares to alternatives, evaluated on Kodak. The additional results are highlighted with blue legend titles in Figure \ref{fig:additional-res}.
First, we consider also applying SGA to the Hyperprior baseline; as shown by the solid blue line in Figure \ref{fig:additional-res}, this also results in a sizable boost in R-D, even larger than what we observe for our more shallow decoders. We hypothesize that this may be caused by a relatively larger inference gap in the Hyperprior architecture than ours with shallow decoders.

Next, we show that simply scaling down existing neural compression models tends to result in worse performance than our approach. We consider two existing architectures: the mean-scale Hyperprior \cite{minnen2018joint} and ELIC \cite{he2022elic}, and slim down their synthesis transforms to match (to our best ability) the FLOPs of our two-layer shallow synthesis. For Hyperprior, we adopt a pruned synthesis transform given by CENIC \cite{johnston2019computationally} (specifically, we use their architecture \# 178, which uses about 7.3 KMACs/pixel, or about 1.4 times of our two-layer synthesis; we keep the hyper synthesis intact). For ELIC, we simply reduce the number of conv channels in the synthesis to be 32, so that it uses about 16.5 KMACs/pixel (we also keep its hyper synthesis intact). We train the resulting architectures from scratch; as shown by the purple (``Johnston 2019 CENIC'') and brown curves (``ELIC XS''), this results in progressively worse R-D performance in the higher-rate regime compared to our two-layer synthesis (red curve). 

Finally, we conduct a preliminary exploration of a JPEG-like architecture for the hyper synthesis transform. We implement this with a single transposed-conv layer with stride 4 and (6, 6) kernels. We applied it on top of our linear JPEG-like synthesis, and observe a 10\% worse BD-rate (green curve $\to$ pink curve in Figure~\ref{fig:additional-res}) but nearly a 10-fold reduction in the hyper synthesis FLOPs (15.18 $\to$ 1.8 KMACs/pixel).

\newpage
\subsubsection{Additional R-D results}\label{sec:additional-rd}
Below we include aggregate R-D results on the 100 test images from Tecnick \cite{tecnick} and 41 images from the professional validation set of CLIC 2018 \footnote{\url{http://clic.compression.cc/2018/challenge/}}. We additionally evaluate on the perceptual distortion LPIPS \cite{zhang2018lpips}.
Overall, we observe that our proposed two-layer synthesis with iterative encoding matches the Hyperprior performance when evaluated on PSNR, but under-performs by $8\% \sim 12 \%$ (in BD-rate) when evaluated on perceptual metrics such as MS-SSIM or LPIPS. This is consistent with results on Kodak in Sec.~\ref{sec:main-experiment-results}. 

\begin{figure}[h]
    \centering
    \includegraphics[scale=0.6]{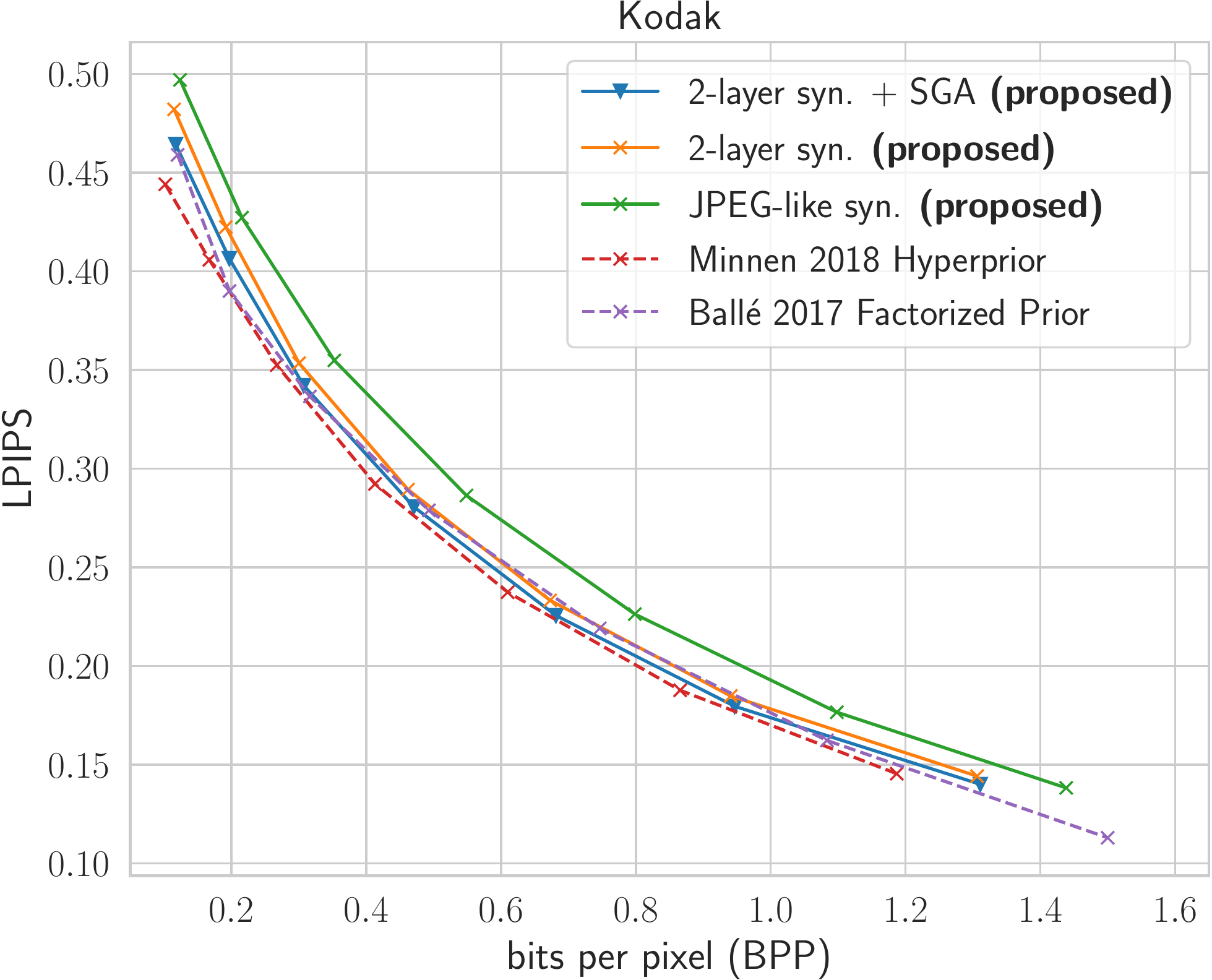}
    \caption{Aggregate LPIPS v.s. BPP performance on Kodak.}
    \label{fig:rd_lpips-kodak}
\end{figure}

\begin{figure}[h]
    \centering
    \includegraphics[scale=0.6]{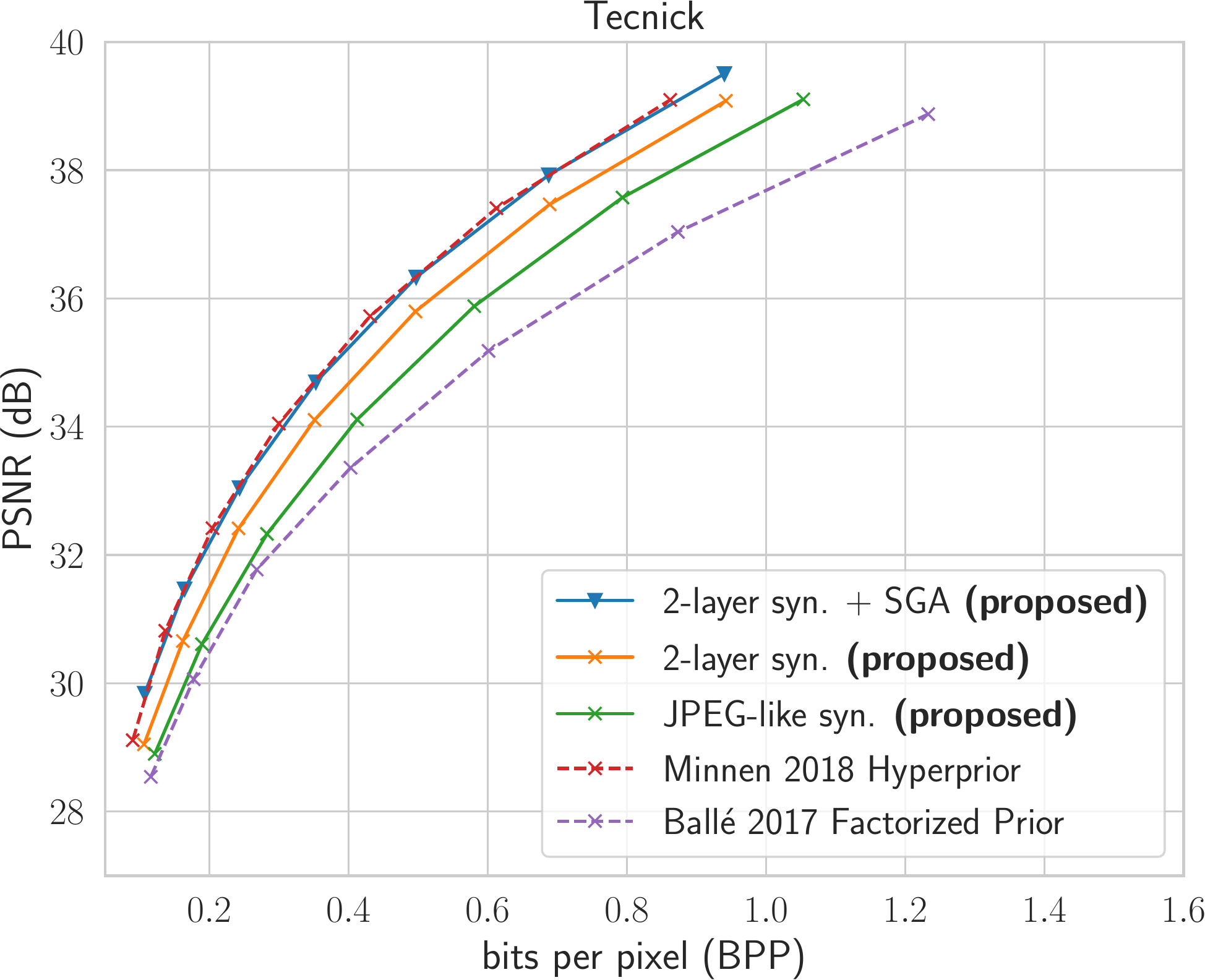}
    \caption{Aggregate PSNR v.s. BPP performance on Tecnick.}
\end{figure}

\begin{figure}[h]
    \centering
    \includegraphics[scale=0.6]{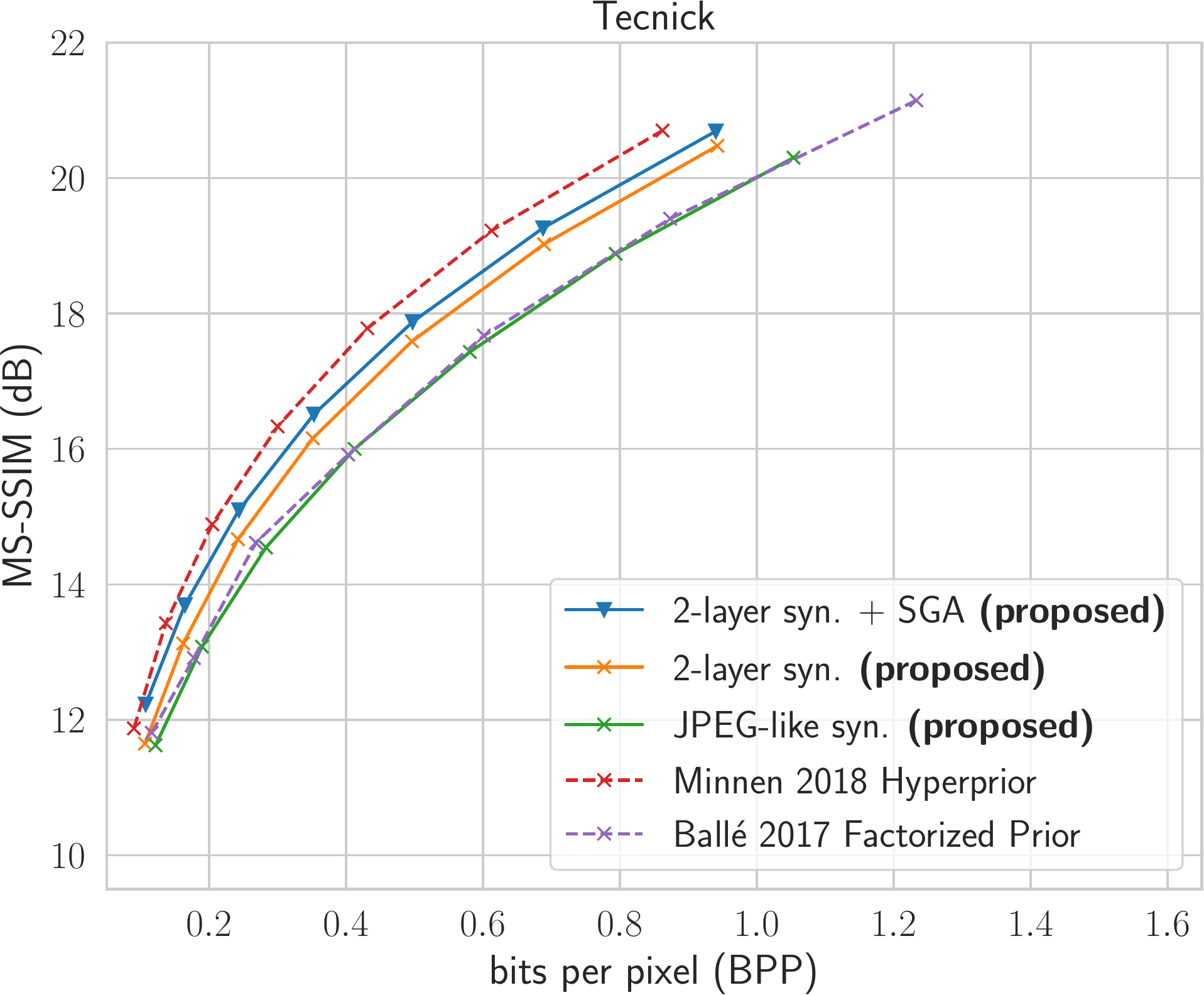}
    \caption{Aggregate MS-SSIM v.s. BPP performance on Tecnick.}
\end{figure}
\begin{figure}[h]
    \centering
    \includegraphics[scale=0.6]{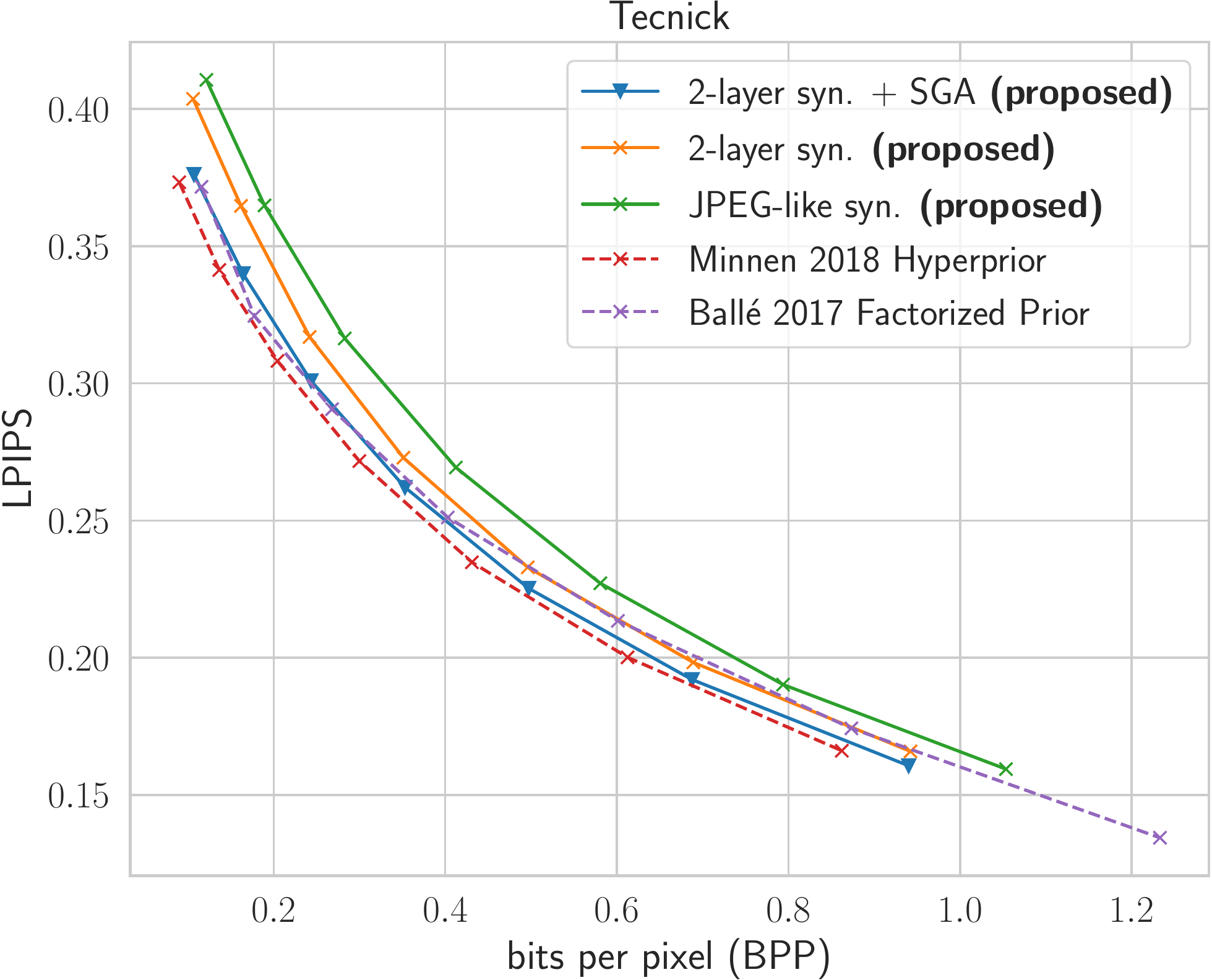}
    \caption{Aggregate LPIPS v.s. BPP performance on Tecnick.}
\end{figure}

\begin{figure}[h]
    \centering
    \includegraphics[scale=0.6]{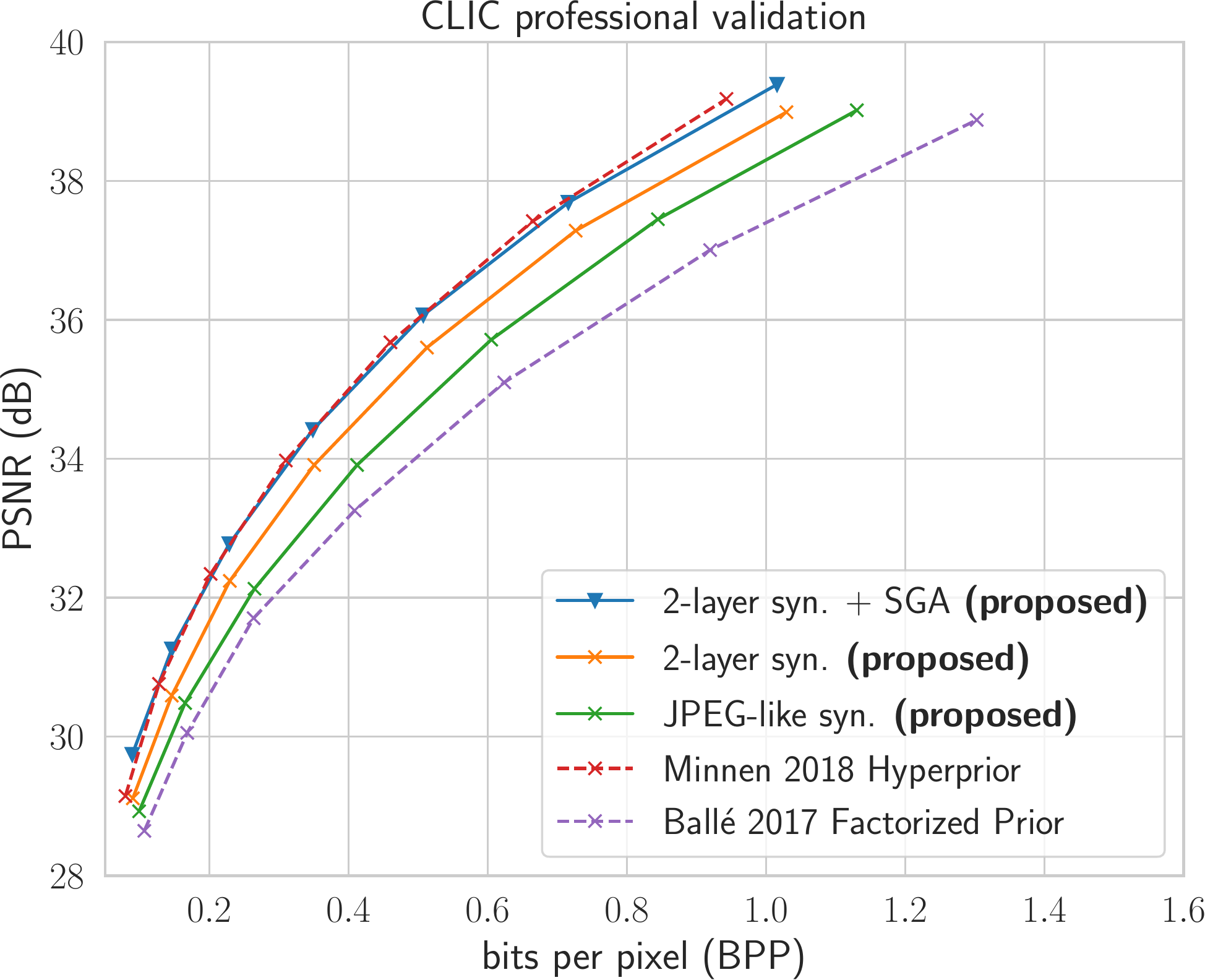}
    \caption{Aggregate PSNR v.s. BPP performance on CLIC professional validation set.}
\end{figure}
\begin{figure}
    \centering
    \includegraphics[scale=0.6]{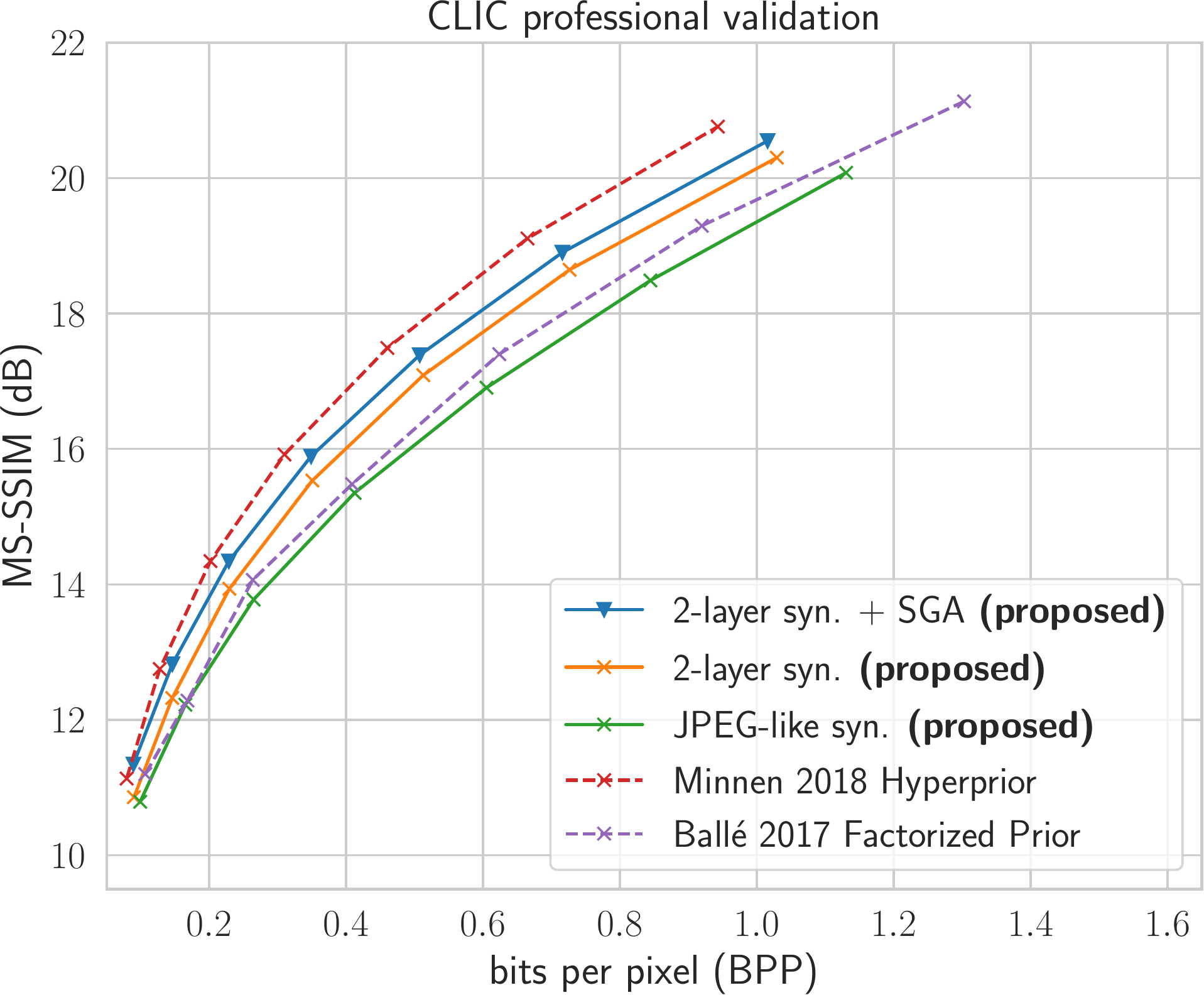}
    \caption{Aggregate MS-SSIM v.s. BPP performance on CLIC professional validation set.}
\end{figure}
\begin{figure}[h]
    \centering
    \includegraphics[scale=0.6]{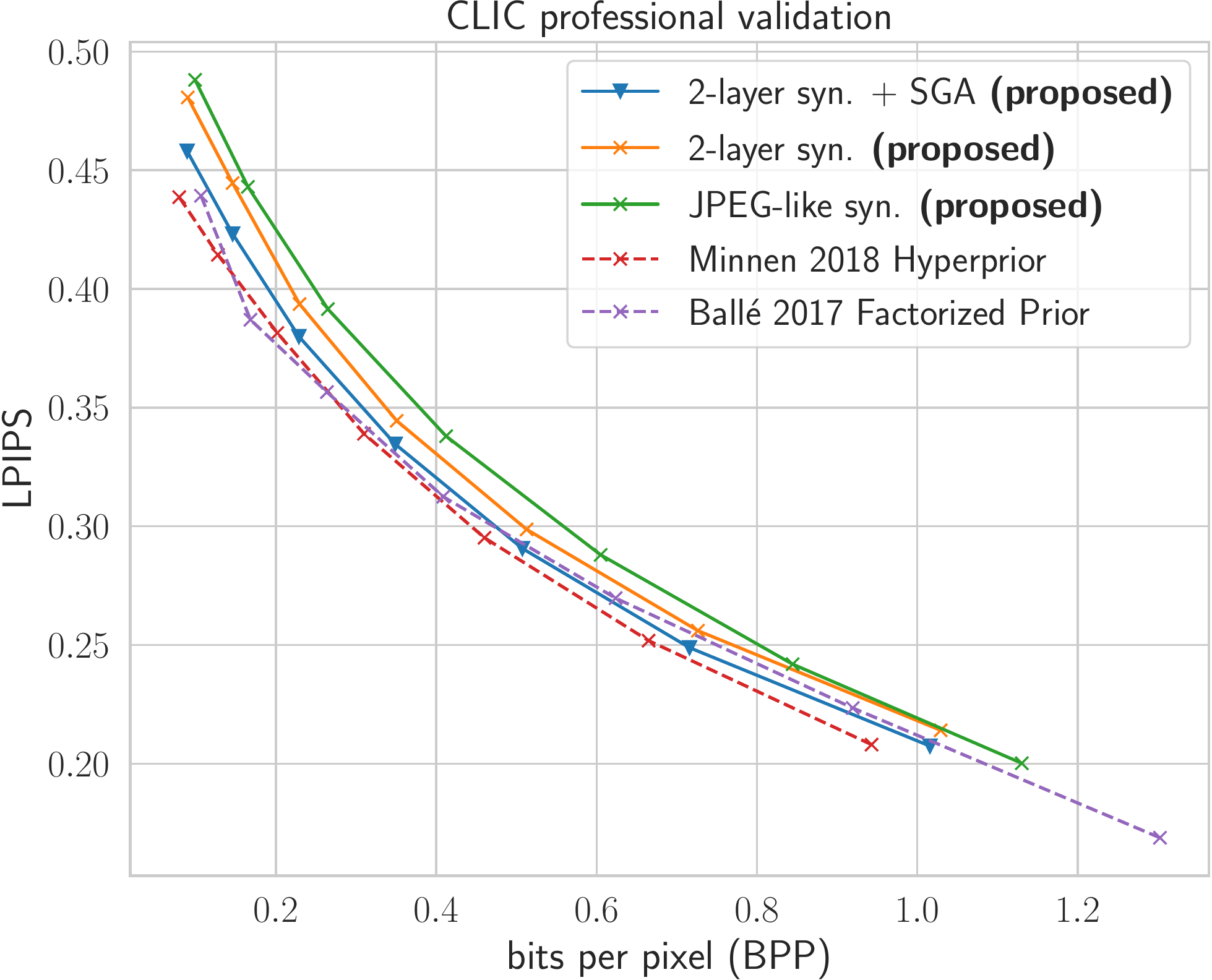}
    \caption{Aggregate LPIPS v.s. BPP performance on CLIC professional validation set.}
\end{figure}

\end{document}